\documentclass[iop]{emulateapj}
\pdfoutput=1 
\usepackage{natbib}
\usepackage[usenames,dvipsnames]{color}
\usepackage{amsmath} 
\usepackage{amssymb}  

\usepackage{booktabs}

\usepackage{hyperref}
\hypersetup{colorlinks=true}

\slugcomment{ACCEPTED FOR PUBLICATION IN PASP, \today}

\shorttitle{Spectroscopic Survey of Eclipsing Binaries with a Low Cost \'{E}chelle Spectrograph}
\shortauthors{Koz\l owski et al.}

\begin{document}

\title{Spectroscopic Survey of Eclipsing Binaries with a Low Cost \'{E}chelle Spectrograph -- Scientific Commissioning}

\author{S.K. Koz\l owski, M. Konacki, P. Sybilski, M. Ratajczak\altaffilmark{1}, R. K. Paw\l aszek}
\affil{Nicolaus Copernicus Astronomical Center}
\affil{Bartycka 18, 00-716 Warsaw, Poland}
\email{stan@ncac.torun.pl}
\and
\author{K.G.~He\l miniak\altaffilmark{2}}
\affil{Subaru Telescope, National Astronomical Observatory of Japan, 650 N. Aohoku Pl., Hilo, HI 96720, USA}

\altaffiltext{1}{Astronomical Institute, Univeristy of Wroc\l aw, Kopernika 11, 51-622 Wroc\l aw, Poland}
\altaffiltext{2}{Subaru Research Fellow}

\begin{abstract}
We present scientific results obtained with a recently commissioned  \'{e}chelle spectrograph on the 0.5-m Solaris-1 telescope in the South African Astronomical Observatory. BACHES is a low-cost slit  \'{e}chelle spectrograph that has a resolution of 21,000 at 5,500 \AA. The described setup is fully remotely operated and partly automated. Custom hardware components have been designed to allow both spectroscopic and photometric observations. The setup is controlled via dedicated software. The throughput of the system allows us to obtain spectra with an average SNR of 22 at 6375 \AA~for a 30-min exposure of a $V=10$ mag target. The stability of the instrument is influenced mainly by the ambient temperature changes. We have obtained radial velocity RMS values for a bright ($V = 5.9$ mag) spectroscopic binary as good as 0.59 km s$^{-1}$ and 1.34 km s$^{-1}$ for a $V = 10.2$ mag eclipsing binary. Radial velocity measurements have been combined with available photometric light curves. We present models of six eclipsing binary systems, and for previously known targets, we compare our results with those available in the literature. Masses of binary components have been determined with 3\% errors for some targets. We confront our results with benchmark values based on measurements from the HARPS and UCLES spectrographs on 4-m class telescopes and find very good agreement. The described setup is very efficient and well suited for a spectroscopic survey. We can now spectroscopically characterize about 300 eclipsing binary stars per year up to 10.2 mag assuming typical weather conditions at SAAO without a single observing trip.
\end{abstract}

\newcommand{\ra}[4]{#1\textsuperscript{h}#2\textsuperscript{m}#3\fs#4}
\newcommand{\de}[4]{#1\degr#2\arcmin#3\farcs#4}


\section{Introduction}
\label{sec:Introduction}
Photometric surveys are well known to produce important scientific discoveries and provide a wealth of data for the astronomical community. For modeling of eclipsing binary stars, however, radial velocity measurements are required to construct more complete models of such systems. Spectroscopic surveys are less numerous than photometric ones due to higher complexity both in software and hardware. At the same time they are technologically appealing, e.g. one of the STELLa Activity telescopes \citep[STELLA;][]{Weber2012}, the Wide Field Spectrograph  \citep[WiFeS;][]{Dopita2010} or Folded Low Order whYte-pupil Double- dispersed Spectrograph  \citep[FLOYDS;][]{Sand2011}. More advanced and technically challenging multi-object spectrographs, such as the Apache Point Observatory Galactic Evolution Experiment \citep[APOGEE;][]{Majewski2015} are optimized for observing dense fields, clusters, galaxies, and conduct all-sky surveys rather than multiple visits to a selected set of targets that are required to model eclipsing binaries. With our hardware setup we extend the capabilities of the Solaris telescope \citep{KozlowskiSPIE2014, Sybilski2014} and make it an adequate tool for studying eclipsing binaries by the means of a spectroscopic survey.

The goal of this new project is to perform a spectroscopic survey of eclipsing binaries from the ASAS\footnote{All Sky Automated Survey - http://www.astrouw.edu.pl/asas/} Catalogue of Variable Stars \citep[ACVS;][]{Pojmanski1997}. In this paper we present results of a preliminary campaign carried out with the Solaris-1 telescope between February and April 2015. The list of targets includes six eclipsing binaries, two spectroscopic standards and two spectroscopic binary stars, all in the 4.9 to 10.2 V magnitude range. Thanks to the very high throughput of the instrument, we were able to show that a simple and inexpensive setup is capable of delivering high quality data. During this campaign we were also able to refine our data reduction pipeline and optimize the configuration of the system. This  will allow us to effectively reduce and analyze data in the future. The second campaign that includes a much larger sample of eclipsing binaries is pending.

The study of eclipsing binaries plays an important role in astrophysics. Thanks to their unique geometrical configuration they can be used to determine fundamental parameters of stars such as radii, masses and effective temperatures, from which other quantities can be derived: bolometric magnitudes, absolute magnitudes, distances, ages. Asteroseismology, theory of stellar structure and evolution, stellar activity, celestial mechanics and other branches of astrophysics can take advantage of precise parameters obtained from the study of binary systems. Although many thousands of eclipsing binary systems are cataloged, only a handful are well described. The most extensive catalog, DEBCat \citep{Southworth2015}, lists 182 systems whose parameters are known with precision high enough (3\%) to test evolutionary codes \citep{Blake2008}. Data in this catalog has been collected since 1975.

Our motivation is to conduct a survey that will significantly increase the count of well described systems. For such a survey access to echelle spectrographs is a serious bottleneck. Additionally, not all eclipsing binaries are equally interesting and their description may not necessarily contribute to the existing database. For example, stars with 1-2 solar masses and radii are well represented in the DEBCat. For these reasons it is important to have a tool that will efficiently and inexpensively allow one to spectroscopically characterize eclipsing binaries and select the most interesting ones for a possible very high precision follow-up with very stable spectrographs such as e.g. HARPS.

In Sec.~\ref{sec:HardwareSetup} we describe our hardware setup, used components and their functionalities, in Sec.~\ref{sec:AcquisitionSoftware} we present our software approach, in Sec.~\ref{sec:SystemPerformance} we discuss the system's performance, throughput, and data quality. The data reduction process is described in Sec.~\ref{sec:DataReduction}. Results for spectroscopic standards, spectroscopic binaries, and eclipsing binaries along with derived models are presented in Sec.~\ref{sec:Results}. We summarize our work in Sec.~\ref{sec:Summary}.

\section{Hardware Setup}
\label{sec:HardwareSetup}
\begin{figure}[ht]
\centering
\includegraphics[width=\columnwidth]{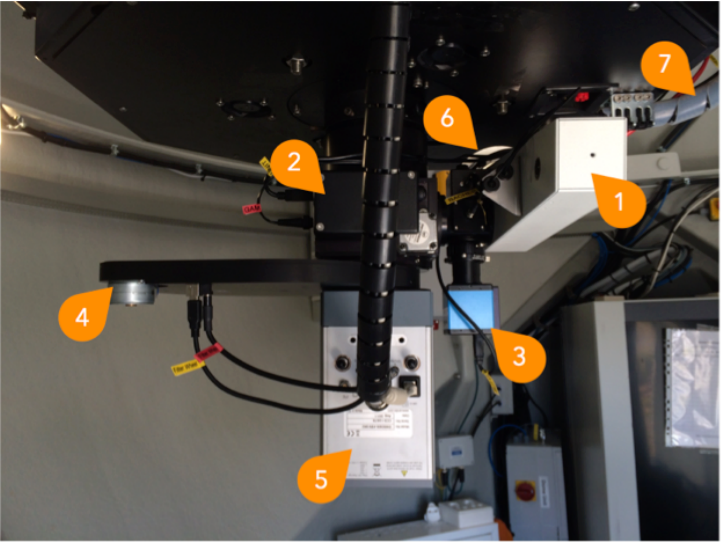}
\caption{Solaris-1 imaging train: 1~--~BACHES spectrograph, 2~--~GAM, 3~--~guide camera, 4~--~filter wheel, 5~--~photometric CCD camera, 6~--~spectroscopic camera (only mounting flange is visible), 7~--~spectrograph control cabling and fiber. }
\label{fig:ImagingTrain}
\end{figure}

The BAsic \'{e}CHEelle Spectrograph (BACHES) has been installed in 2014 on the Solaris-1 telescope in the South African Astronomical Observatory that is part of a network of robotic telescopes. An early prototype of this instrument has been briefly tested on the Solaris-4 telescope in Argentina \citep{Kozlowski2014}. The currently used spectrograph is an improved production version with an increased resolution and mechanical enhancements. The location has been chosen due to the fact that two Solaris telescopes are located at this site. Carrying out spectroscopic observations on one of the telescopes does not interfere with the global photometric coverage that is an important feature of the entire network. Solaris-1 is a Ritchey-Cretien f/15 0.5-m telescope on a DDM-160 direct drive German Equatorial Mount. Both the telescope and the mount have been supplied by Astrosysteme Austria (ASA). Originally the imaging train consisted of a field rotator, filter wheel, and CCD camera. A custom built guide and acquisition module (GAM) has been installed in place of the field rotator to accommodate the spectrograph. Leaving the field rotator was impossible due to the limiting back-focus distance (Fig.~\ref{fig:ImagingTrain}). The GAM allows spectroscopic and photometric observations to be conducted without the need of changing the hardware setup of the imaging train, i.e. permits remote and automatic observations in both modes. The GAM consists of a moving mirror housed in an aluminum chassis,  simple limit switch, stepper motor directly coupled to the mirror's rotating axis via a nylon coupling, and a micro controller-based motor controller. The mirror mechanism is additionally spring-loaded to ensure stable end-positions without the need of powering the motor when it is stationary to avoid heat-induced turbulences in the optical path. The limit switch is used for position calibration, as the motor is not equipped with encoders. A simple open-loop control approach is sufficient for this task. The GAM has two optical output ports. The straight-through one is designed for the photometric CCD camera and filter wheel assembly, i.e. the original observing mode of the telescope and has a standard 4" ASA dovetail interface. The second port is designed for the BACHES spectrograph. The device meets our expectations in terms of mechanical stiffness - GAM structural analysis revealed that the GAM itself introduces shifts to the photometric camera's CCD in the order of the size of the CCD's pixels, which is less than the camera-fliter wheel assembly interface.

%
%
\section{Acquisition Software}
\label{sec:AcquisitionSoftware}

Custom software components have been written specifically for the spectroscopic mode. The control software runs in parallel with the observatory management software suit and will become part of it in the future allowing fully automated spectroscopic data acquisition. Hardware components directly involved in the spectroscopic observation flow include: BACHES, GAM, slit-view camera, spectroscopic camera, Remote Calibration Unit (RCU), telescope (including focuser). Features of these components are presented in Tab. \ref{tab:components}. Data presented in this paper has been acquired in a semi-automatic way, i.e. the observer had to supervise the observatory using a remote desktop connection from Torun, Poland. Tasks that have been automated include slewing to a target selected from a predefined list, repositioning and centering on-slit, calibration frame acquisition -- thorium-argon (ThAr hereafter) and quartz lamps, science frame acquisition and guiding. Necessary star detection, astrometric and guiding algorithms have been implemented in the control software. All software components running the observatory regardless of the operating mode (photometry, spectroscopy) are now custom and optimized for the described setup. Yet, thanks to a modern approach, they can be easily modified and reused in other projects. 

\begin{table*}
\caption{List of spectroscopic mode components and hardware/software implementation details.}
\begin{center}
\scriptsize{
\begin{tabular}{llll}
\hline
\hline
Component & Description & Interface & Driver \\ 
\hline
GAM & custom design & USB 2.0 & custom protocol, .\textsc{net} dll\\
slit-view camera &  The Imaging Source DMK 21AU04.AS, & USB 2.0 & wrapper around TIS .\textsc{net} dll\\
		          & 640x480 5.6 $\mu m$ square pixels & & \\
spectroscopic camera & Finger Lakes Instrumentation & USB 2.0 & .\textsc{net} wrapper around FLI C dll \\
				  & ML1306,  1536x1024 9.0 $\mu m$ square pixels & & \\
RCU & Baader Planetarium, control of ThAr & Ethernet & custom .\textsc{net} library\\
and Quarz lamps, flp-mirror & & \\
telescope & ASA DDM-160 & USB 2.0 & ASCOM\\
\hline
\end{tabular}
}
\end{center}
\label{tab:components}
\end{table*}

\section{System Performance}
\label{sec:SystemPerformance}
\subsection{Spectral Resolution, Data Quality and Throughput}

\begin{figure}
\includegraphics[width=\columnwidth]{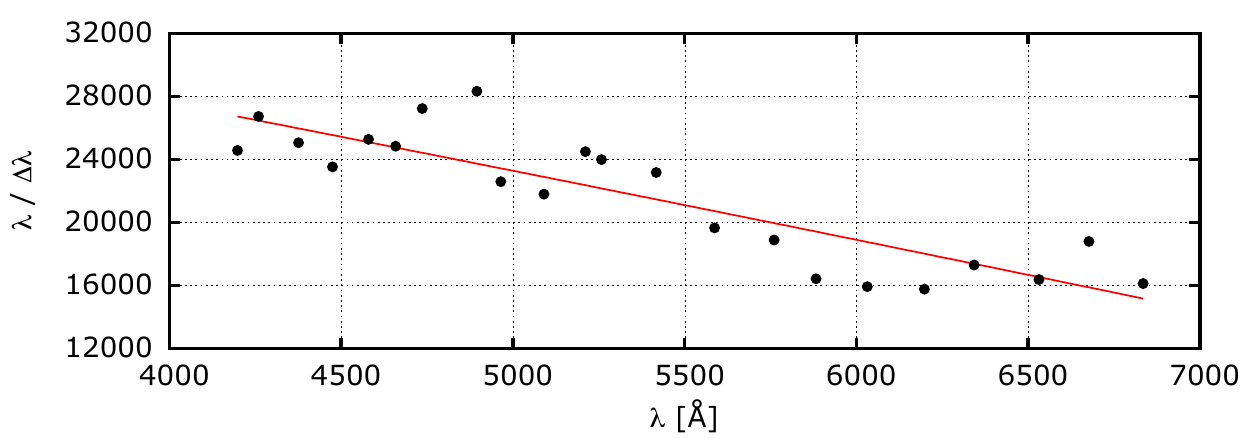}
\caption{BACHES resolution as a function of wavelength obtained from ThAr calibration spectra. A 2nd order polynomial fit (solid line) and measured values (dots) are plotted.}
\label{fig:Resolution}
\end{figure}

\begin{figure}
\includegraphics[width=\columnwidth]{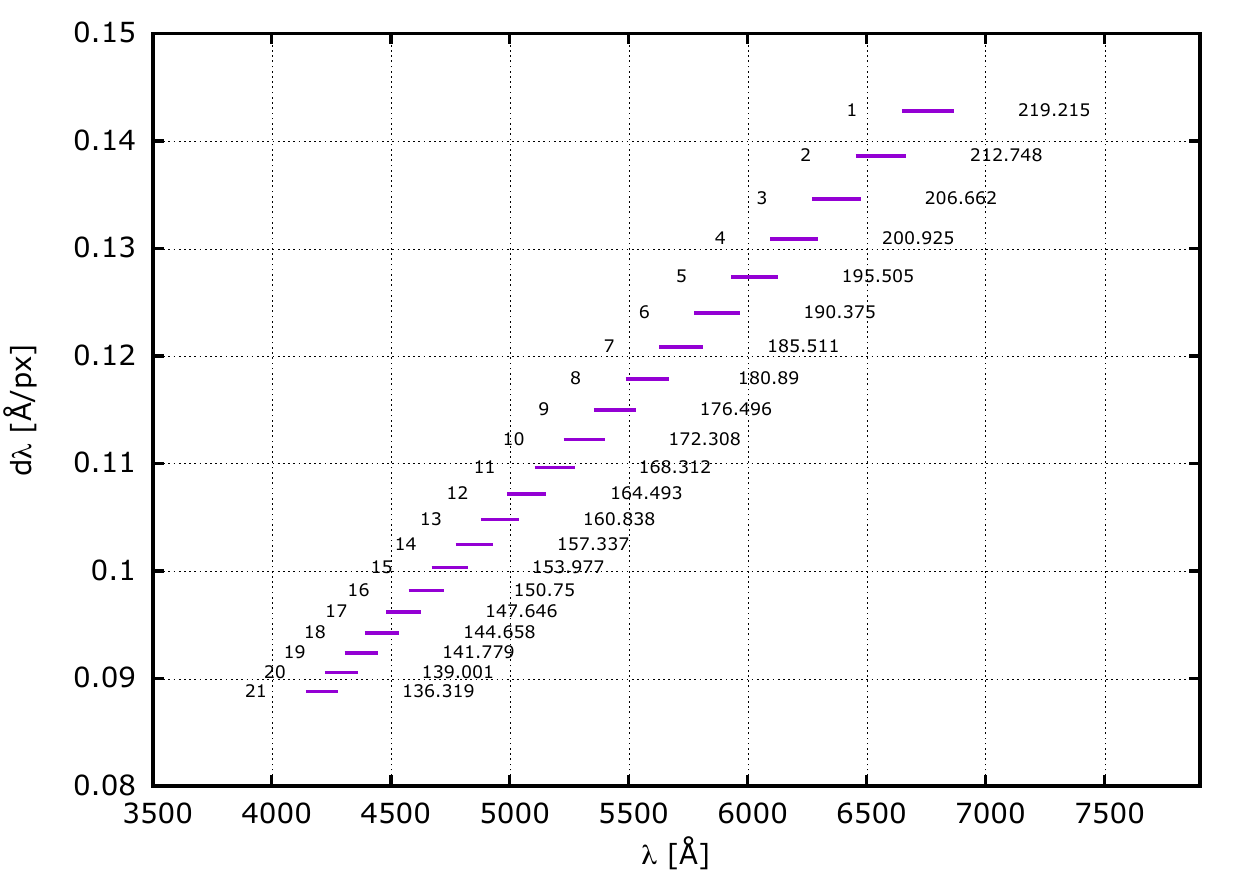}
\caption{BACHES orders configuration. For each order the lines show average pixel scales as a function of wavelength, and correspond to the wavelength ranges. Orders are numbered from 1-21. Their lengths (in \AA) are also given. }
\label{fig:Orders}
\end{figure}

The spectral resolution of BACHES varies with wavelength and exceeds 26,000 on the blue side of the spectrum and drops down to 15,000 on the red side of the spectrum (Fig. \ref{fig:Resolution}). While the echellogram in the described configuration consists of 26 orders, our data reduction pipeline uses only the 21 best exposed orders which cover a range from 4143.7 to 6648.4 \AA. The width of orders in wavelength varies from 136.3 \AA~to 219.2 \AA~and the orders overlap by 18 to 57 \AA~on each side. Wavelength scale varies from 0.089 to 0.143 \AA/px (Fig. \ref{fig:Orders}).

With the current semi-automatic setup, the overhead per each 30 min exposure is  $\sim$4.5 min. This overhead includes telescope slewing, slit centering and ThAr frames taken before and after each science exposure but does not include a set of flat field (quartz lamp) frames that are taken once per observing night. Although the efficiency is satisfactory, it can be improved by adopting a fully automatic approach. 

Data quality and real-life throughput are determined by a combination of factors that include seeing, wind speed, and focus quality. The telescope mount in a fully open clamshell dome can correct for wind-induced disturbances as long as the wind is stable and its speed does not exceed 25 km h$^{-1}$. If it does and/or is gusty, partially closing the dome windward side usually helps and allows for good quality of observations at wind speeds up to 35 km h$^{-1}$. Above that value only bright targets should be considered. 

\begin{figure}
\includegraphics[width=\columnwidth]{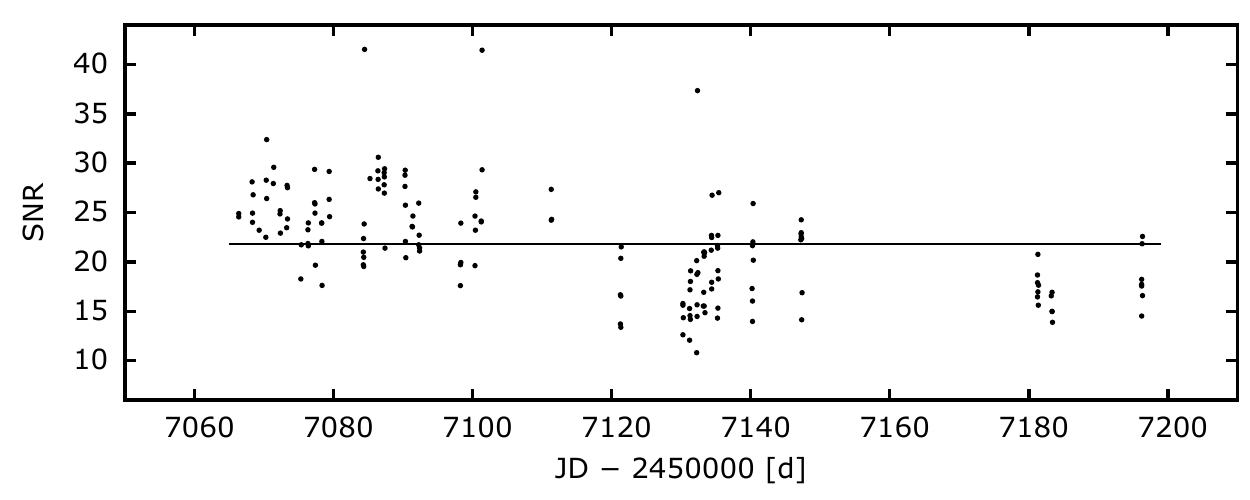}
\caption{Computed SNR values of all acquired object spectra normalized to a 30 minute exposure of a V=10 mag target at 6375\AA. The horizontal line represents an average value of 21.78. It should be noted, however, that a linear, negative trend is visible in the plot. It is caused by the fact that the program stars were observed at roughly the same time in the same order throughout the campaign. Due to the time span of the observations zenith distances would increase with time for most objects, causing a decrease in the SNR values.}
\label{fig:snr}
\end{figure}

The average SNR value at the brightest part of the spectrum, 6375\AA, for a 30 minute exposure of a V=10 mag target is 21.78 (Fig. \ref{fig:snr}). During good seeing conditions and calm air it is realistic to expect and SNR of 30 for the described case. 

\begin{figure}
\includegraphics[width=\columnwidth]{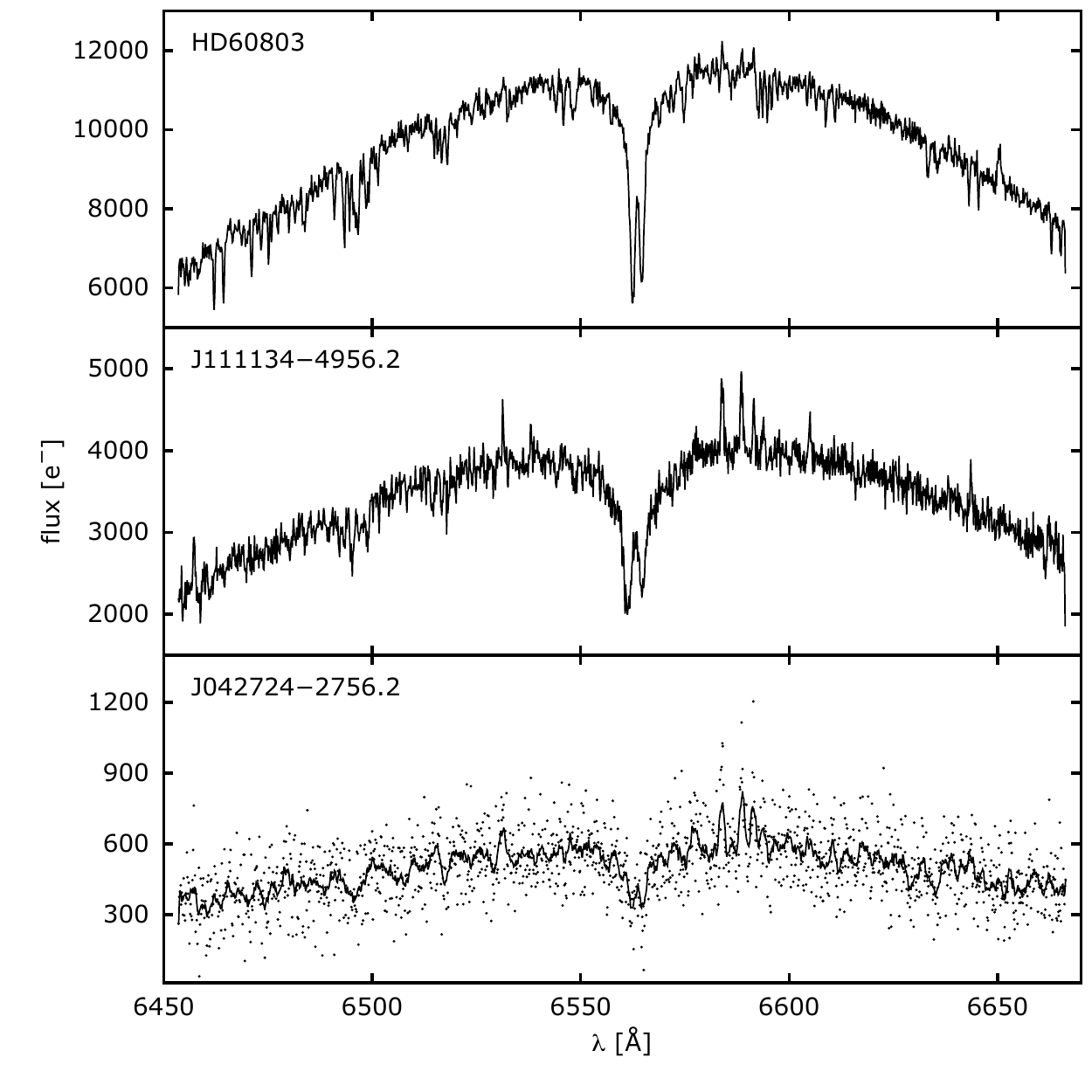}
\caption{An \'{e}chelle order ($n = 2$) from BACHES for three binary systems HD60803, J111134-4956.2 and J042724-2756.2. The epoch of the spectra is such that the differences between values of radial velocities of the components of the binaries are relatively large and one can see the H$\alpha$ from both components. The dots in the lowest panel represent the actual recored flux of J042724-2756.2. The solid line represents a smoothed spectrum (boxcar smoothing with 10 pixels) that clearly reveals double H$\alpha$  lines. }
\label{fig:spectra}
\end{figure}

Sample spectra of three targets of different magnitudes are shown in Fig. \ref{fig:spectra}. 

\subsection{Instrument Stability}

\begin{figure*}
\includegraphics[width=\textwidth]{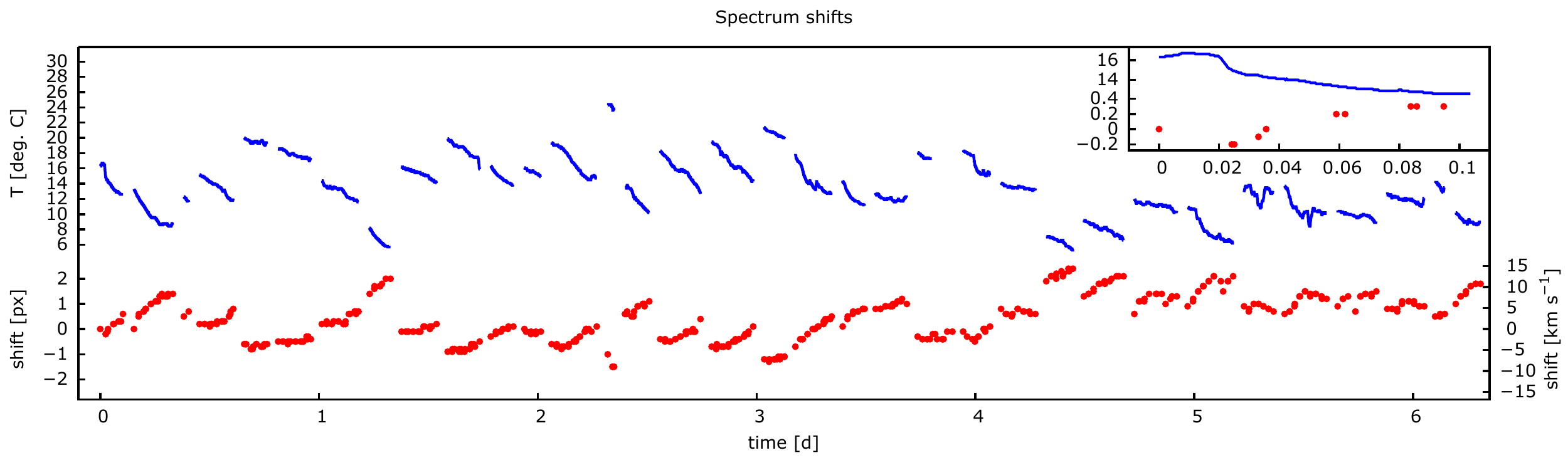}
\caption{ThAr shifts in the dispersion direction with respect to the first acquired calibration frame during the campaign. Ambient temperature recorded with the external weather station and shifts in pixels and corresponding radial velocity values are plotted against observation time. Time gaps between consecutive observing sessions have been removed to better visualize the stability of the instrument. 2.5 observing hours are shown inset. Units are as in the main plot. It is clearly visible how the instrument reacts to the temperature change \textit{bump} with slight delay.}
\label{fig:Shifts}
\end{figure*}

In the described setup the spectrograph is subject to varying atmospheric conditions throughout the observing session. During a typical observing session from dawn till midnight local time an ambient temperature drop of up to 5\arcdeg C was observed. This corresponds to 1.5 pixel shift in the dispersion direction what translates to ~10 km s$^{-1}$ in radial velocity (RV). Fortunately, the temperature change is much smaller during a single exposure and can be successfully accounted for using pre- and post-exposure ThAr calibration spectra. Figure \ref{fig:Shifts} shows details of measured spectrum shifts during the campaign. 

\section{Data Reduction and Analysis}
\label{sec:DataReduction}
\subsection{Data Reduction}
During the campaign data has been acquired according to the following loop: centering target on slit, ThAr spectrum, object spectrum, ThAr spectrum, slew to the next target. A set of bias and flat frames has been acquired, depending on the weather conditions, before or after the observing session. Flat frames are quartz lamp spectra, no sky flats nor darks have been taken. The data reduction process uses \textsc{iraf} tasks with custom scripts and enhancements and involves several steps. (1) Master bias and master flat frames are created. Two types of master flats are generated -- one being a standard averaged flat used for order tracing and the other a median filter smoothened and normalized frame for calibrating pixel-to-pixel response of the camera. ThAr and object frames are divided by this type of flat. (2) Cosmic rays are removed from object frames using an \textsc{iraf} implementation of the \textsc{l.a.cosmic} algorithm by \cite{Dokkum2012}. (3) The master flat frame is used to trace orders. Spectra are extracted from ThAr and object frames. (4) Wavelength calibration is performed on ThAr spectra and the mean of the two (before and after) ThAr wavelength solutions is applied to object spectrum. Barycentric corrections are computed and added to the frames' headers for use in subsequent steps.

\subsubsection{Additional Wavelength Calibration}

\begin{figure}
\begin{center}
\includegraphics[width=0.5\textwidth]{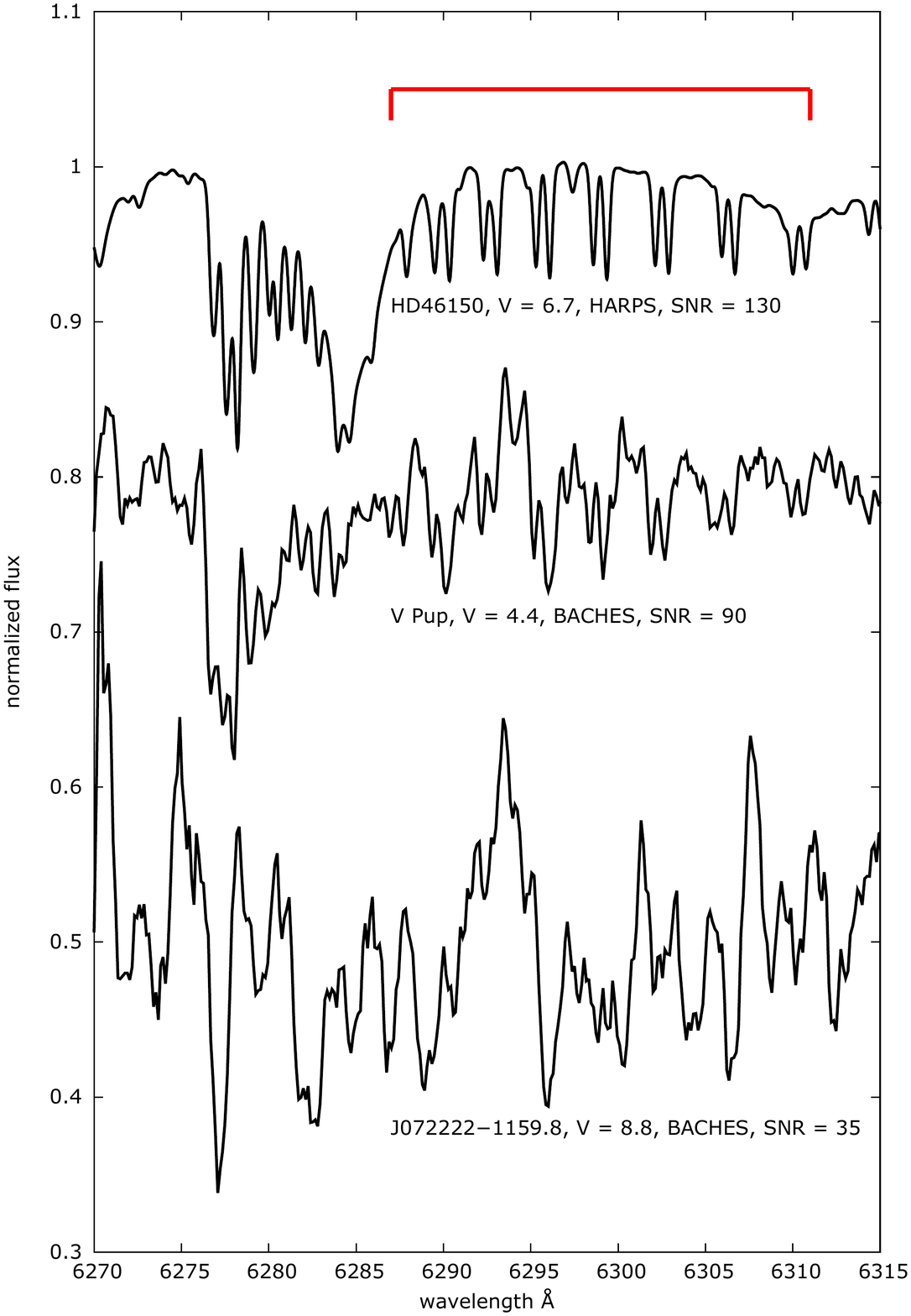}
\caption{
O$_2$ $\gamma$-band portion of three normalized spectra (region of interest is marked with the red bracket). The HARPS spectrum has been additionally convolved with a PSF corresponding to BACHES' resolution to widen and shallow the absorption lines for better comparison with the remaining data. The plot illustrates how strongly the SNR of the spectrum influences the visibility of telluric lines. These are clearly visible in HD46150. For V Pup, the lines are still present though the reduced SNR makes them less obvious. In the last plot, telluric lines disappear in the noise.}
\label{fig:Telluric}
\end{center}
\end{figure}

As described above, spectra have been calibrated using reference ThAr spectra taken before and after each science exposure. Telluric lines calibration  can be used in conjunction with the ThAr-based wavelength solution. Various molecules present in the Earth's atmosphere leave a footprint on the spectrum of the stellar light before it reaches the spectrograph. These absorption lines are also present in the optical wavelength range and can be used to correct the wavelength solution for object spectra. Although the format of BACHES' echellogram has been adjusted to avoid contamination of the spectrum by telluric lines, some O$_2$ ($\gamma$-band) and water vapor absorption lines fall within the range of the instrument. These lines, however, are not easy to  use as a robust calibration method. Two main factors determine this: (a) the absolute (expressed in counts) depth of telluric lines strongly depends on the amount of light that travels through the atmosphere, i.e. the better the SNR, the deeper and more useful are the lines (the relative depth of the lines depends on airmass and water vapor content but for a low SNR spectrum these lines may be lost in noise); (b) telluric lines are very narrow and using a small to medium resolution spectrograph causes their recorded depth to decrease. Figure \ref{fig:Telluric} shows a comparison of the $\gamma$-band portion of the spectrum obtained for three different targets: HD46150, a V = 6.7 O-type star using HARPS \footnote{Data obtained from the ESO archive.}, V Pup, a V = 4.4 mag binary with B-type components using BACHES and J072222-1159.8, a V=8.8 eclipsing binary that is one of the campaign stars. Water vapor lines, e.g. in the 6490.7910 - 6574.8491 \AA ~range are present in BACHES spectra as well. These lines may also be used to correct the wavelength solution. We plan to adopt this approach in future work having in mind that for late-type spectra of sharp-lined stars the telluric lines can often be blended with stellar lines, so cross-correlation (3 dimensional) or disentangling might be necessary in such cases.

\subsection{Radial velocity calculation}
In order to determine initial values of RVs our own implementation of two dimensional cross-correlation technique \textsc{todcor} \citep{Mazeh1994, Zucker2003} is performed. It uses various synthetic spectra computed with the \citet{kur92} \textsc{atlas9} and \textsc{atlas12} codes as references. The template spectra span 3500 to 7000 \AA, hence cover the entire BACHES spectral range and have been chosen to correspond to the parameters of the binary system's components and the spectrograph resolution. It should be noted, however, that at this level of RV precision a precise selection of the template spectra is not a crucial factor. One-dimensional crosscorrelation functions are computed in \textsc{iraf} per order, for the wavelength range that is defined by that order.  Bootstrap analysis of TODCOR correlation maps created by adding randomly selected single-order maps is carried out to compute the formal RV errors. To avoid errors underestimation and obtain the best fit (reduced $\chi^2$ $\approx$ 1), formal errors are increased in quadrature.

\subsection{Data Analysis}
RV and photometric (V-band light curves from ACVS) data analysis is performed using following codes: our own \textsc{v2fit} which fits a double-Keplerian RV orbit, \textsc{jktebop}  \citep{Southworth2004a,Southworth2004b} which deals with light curves (LC) modeling (for eclipsing binaries), \textsc{phoebe} \citep{Prsa2005} to model physical properties of stars, and \textsc{jktabsdim} \citep{Southworth2004a,Southworth2004b} in order to obtain absolute values and uncertainties of stellar parameters. In our analysis process $T_0$ is defined as in \textsc{phoebe} \citep{Helminiak2009} and the primary component is adopted as a star being eclipsed during the primary eclipse. Details of modeling procedure are described in \cite{Ratajczak2013mnras} and contain the use of above mentioned codes to minimize the $\chi^2$ function with Lavenberg-Marquadt algorithm (\textsc{v2fit}), fit a geometrical model to LC of eclipsing binaries (\textsc{jktebop}), and determine the values of systems components parameters (\textsc{phoebe}, \textsc{jktabsdim}).
\begin{figure*}
\includegraphics[angle=-90,width=0.32\textwidth]{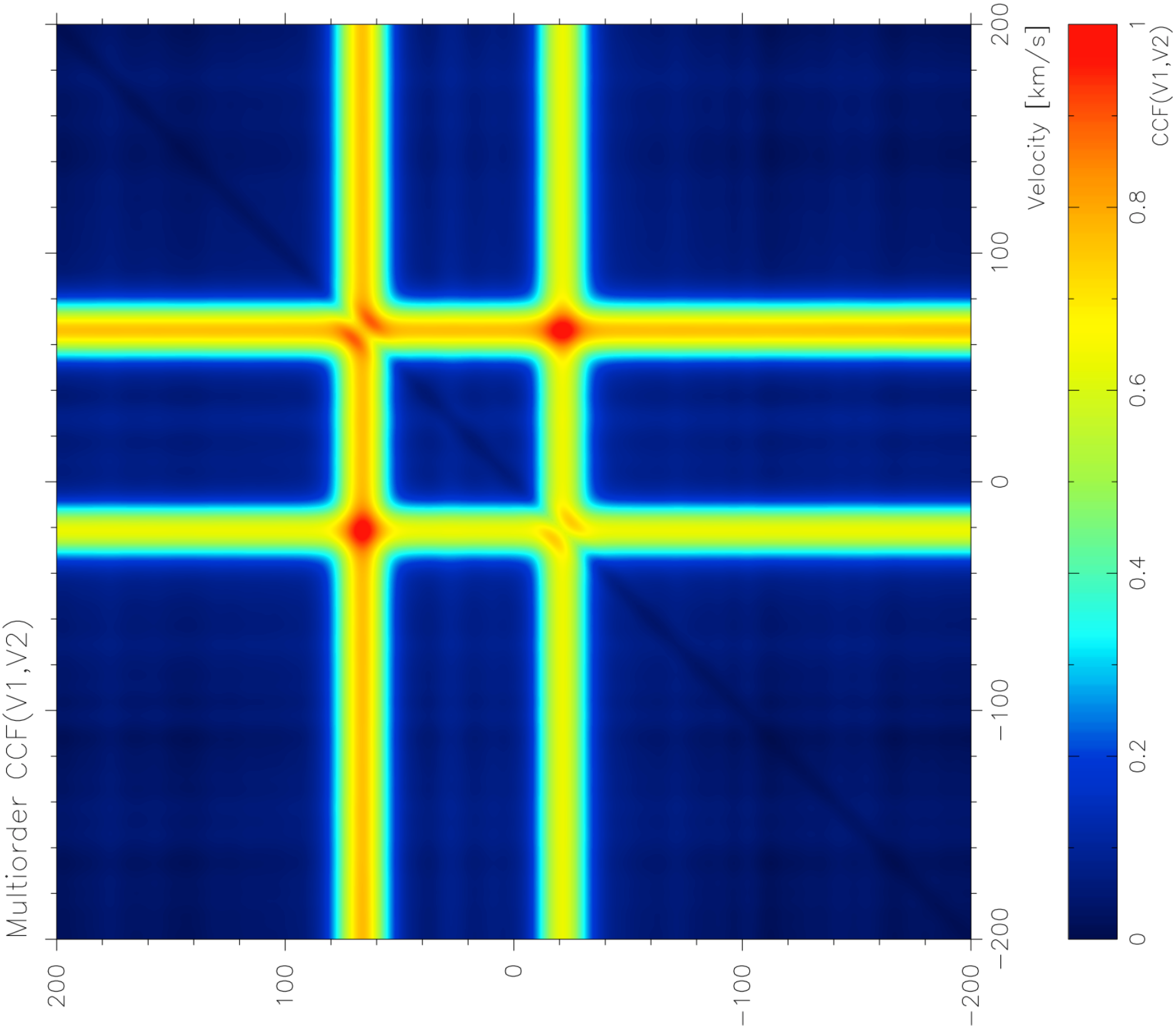}
\includegraphics[angle=-90,width=0.32\textwidth]{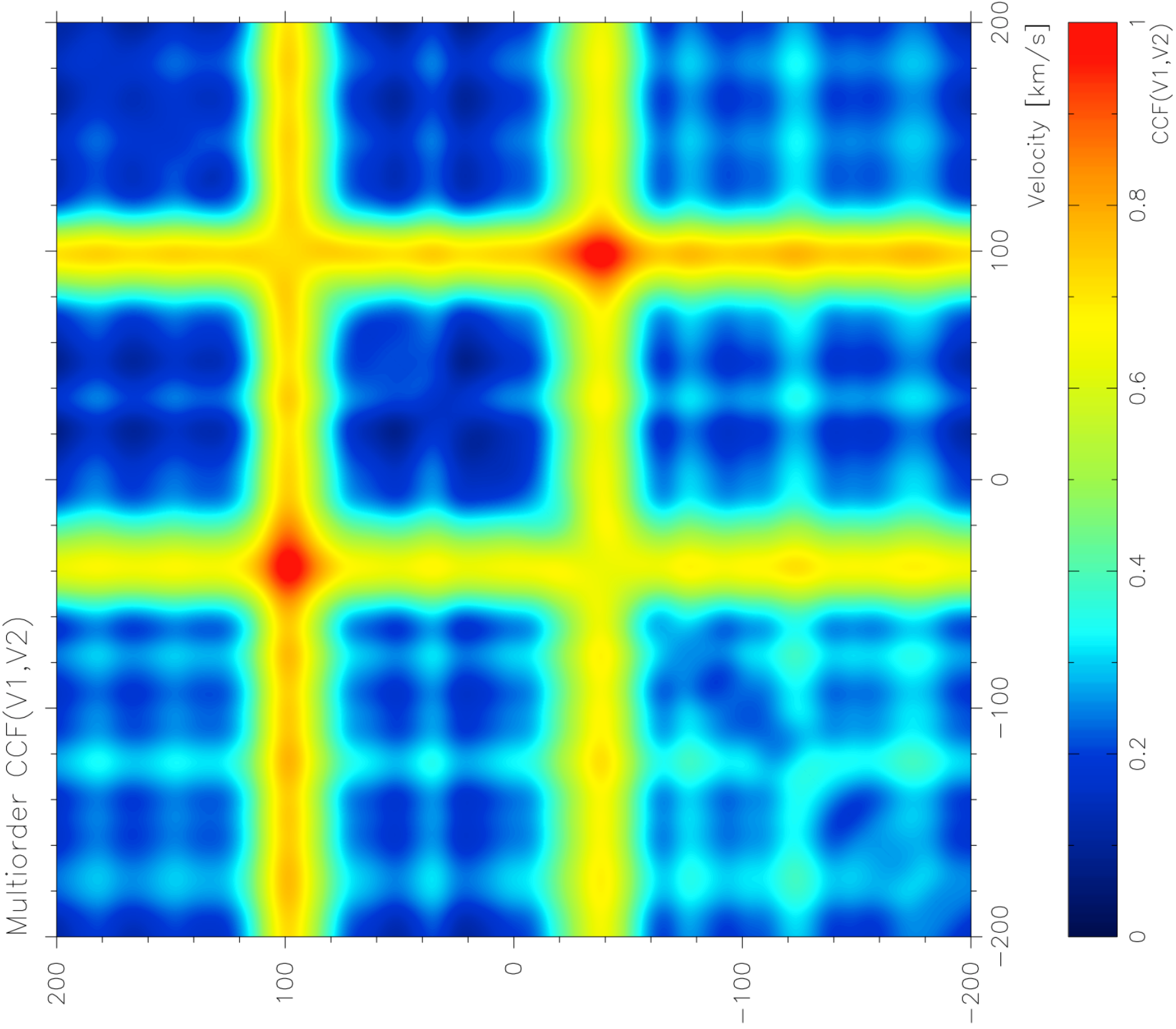}
\includegraphics[angle=-90,width=0.32\textwidth]{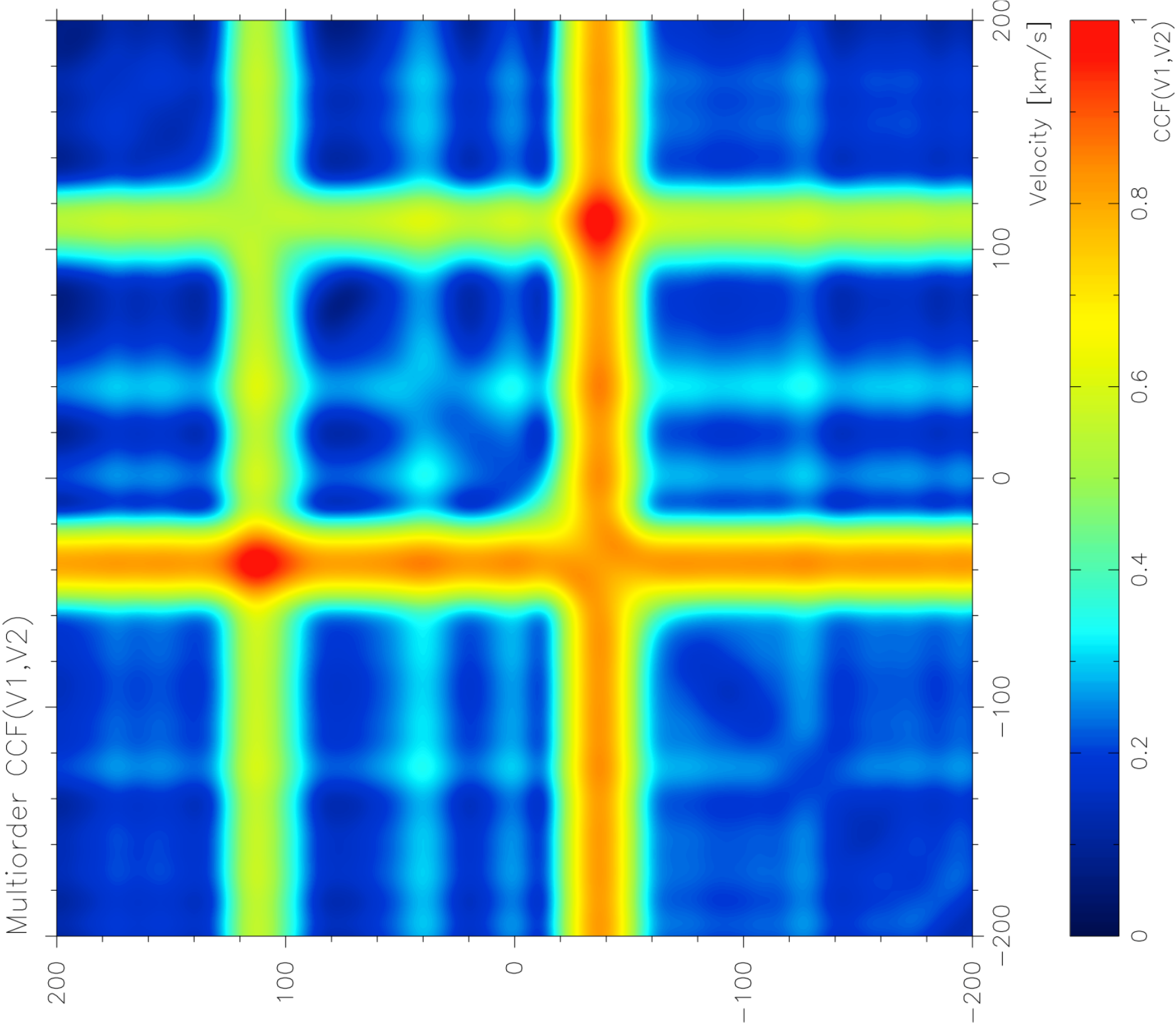}
\caption{Todcor 2D cross-corellation multiorder maps \citep{Zucker2003} for three objects: HD60803, a bright V=5.9 mag binary (left panel); J071626+0548.8, faintest object in the campaign, V = 10.2 mag (middle panel); J072222-1159.8, V=8.8 mag binary system consisting of two components having large luminosity difference -- the signal from the more luminous component is much stronger than that of the less luminous one what is manifested by the cross-correlation function values at the horizontal and vertical line pairs (right panel). }
\label{fig:Todcor}
\end{figure*}

\section{Program Stars}
\label{sec:Results}

Program stars are listed in Tab. \ref{tab:ProgramStars}. The following sections describe the results obtained for these targets.

\begin{table*}
\caption{Program stars. EB - eclipsing binary, STD - spectroscopic standard, SB - spectroscopic binary.}
\begin{center}
\begin{tabular}{cccccccc}
\hline
\hline
ID 				& R.A. (J2000) 	& Dec. (J2000) & V(mag)  & exp. time (s) 	& spectra & group & type \\ 
\hline
J042724-2756.2	& \ra{04}{27}{24}{0} 	& \de{--27}{56}{12}{0} 	& 9.9		& 1800		& 16 & EB	  & F5 \\
J061212-1215.8	& \ra{06}{12}{11}{9} 	& \de{--12}{15}{47}{9} 	& 8.8		& 1800		& 16 & EB	  & F3 \\
HD45184			& \ra{06}{24}{43}{9} 	& \de{--28}{46}{48}{4} 	& 6.4		& 600			& 18 & STD & G1	  \\
J071626+0548.8	& \ra{07}{16}{26}{0} 	& \de{+05}{48}{47}{9} 	&10.2	& 1800		& 16 & EB	 & G6  \\
J072222-1159.8	& \ra{07}{22}{22}{0} 	& \de{--11}{59}{48}{0} 	& 8.8		& 1800		& 16 & EB	  & A0 \\
HD60803			& \ra{07}{36}{34}{7} 	& \de{+05}{51}{43}{8} 	& 5.9		& 600			& 16 & SB	 & G0 \\
HD87810			& \ra{10}{07}{07}{6}	& \de{--21}{15}{20}{6}	& 6.7		&  900			& 14 & SB	  & F3\\
J111134-4956.2	& \ra{11}{11}{34}{0}	& \de{--49}{56}{12}{0}	& 8.3		& 1800		& 12 & EB	 & A3 \\
HD102365		& \ra{11}{46}{31}{1}	& \de{--40}{30}{01}{3}	& 4.9		& 400			& 12 & STD & G2	  \\
J155259-6637.8	& \ra{15}{52}{59}{0}	& \de{--66}{37}{47}{9}	& 9.0		& 1800		& 10 & EB	  & F3\\

\hline
\end{tabular}
\end{center}
\label{tab:ProgramStars}
\end{table*}

\subsection{Spectroscopic Standards}

Two stars -- HD45184 (G1V) and HD102365 (G2V) - have been chosen to verify the stability of the spectrograph and have been observed throughout the campaign. Both have planetary companions, hence are well studied in terms of radial velocity measurements: HD45184 has a planet in 5.9 day orbit and the corresponding RV amplitude is 4.77 ms$^{-1}$ \citep{Mayor2011} and the systemic velocity is -3.7584 $\pm$ 0.0008 (HARPS). HD102365 has a planet in a 122 day orbit and the corresponding RV amplitude is 2.40 ms$^{-1}$ \citep{Tinney2011}. Based on the available RV data from HARPS \citep{Zechmeister2012} we have computed the mean RV equal 17.011 $\pm$ 0.003 kms$^{-1}$ for this system. The RV amplitudes caused by planetary companions are much too low to be detectable by our instrument, making these stars good standards for our commissioning purposes. For HD45184 we have measured RV= -4.58 $\pm$ 0.25 kms$^{-1}$ with an RMS\footnote{In this case the RMS represents the scatter around the RV model which is a straight line fit to the data.} value of 1.00 kms$^{-1}$ and for HD102365 RV= 17.4 $\pm$ 0.4 kms$^{-1}$ with a RMS value of 1.32 km s-1. These results are consistent with values available in the literature. RV plots are shown in Fig. 8. Even though absolute systemic velocities computed using BACHES might slightly differ from values obtained with HARPS, for RV modeling of eclipsing binaries we are mostly interested in relative velocities. This statement holds as long as we use the same instrument throughout the survey.  

\begin{figure}
\includegraphics[width=0.48\textwidth]{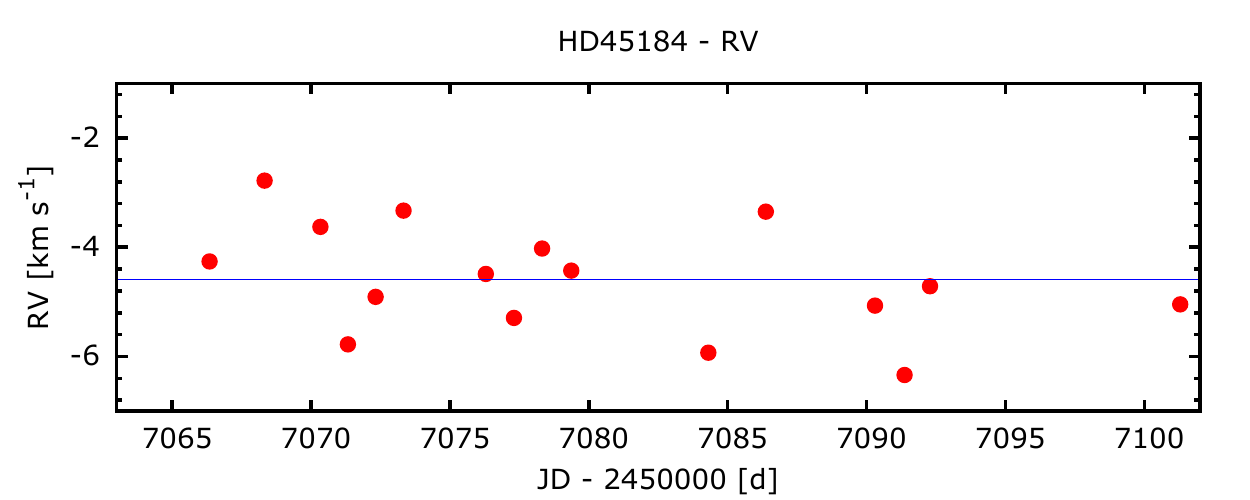}
\includegraphics[width=0.48\textwidth]{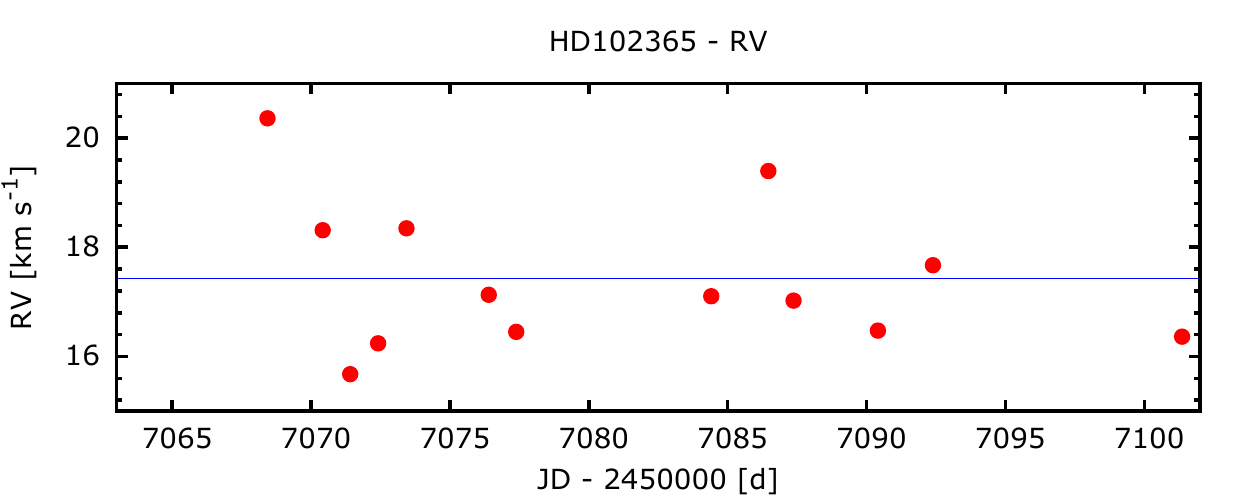}
\caption{RV measurements for spectroscopic standards: HD45184 and HD102365. Solid lines indicate constant function fits.}
\label{fig:plotHD45184}
\end{figure}

\subsection{Spectroscopic Binaries}

\begin{figure*}
\includegraphics[width=0.48\textwidth]{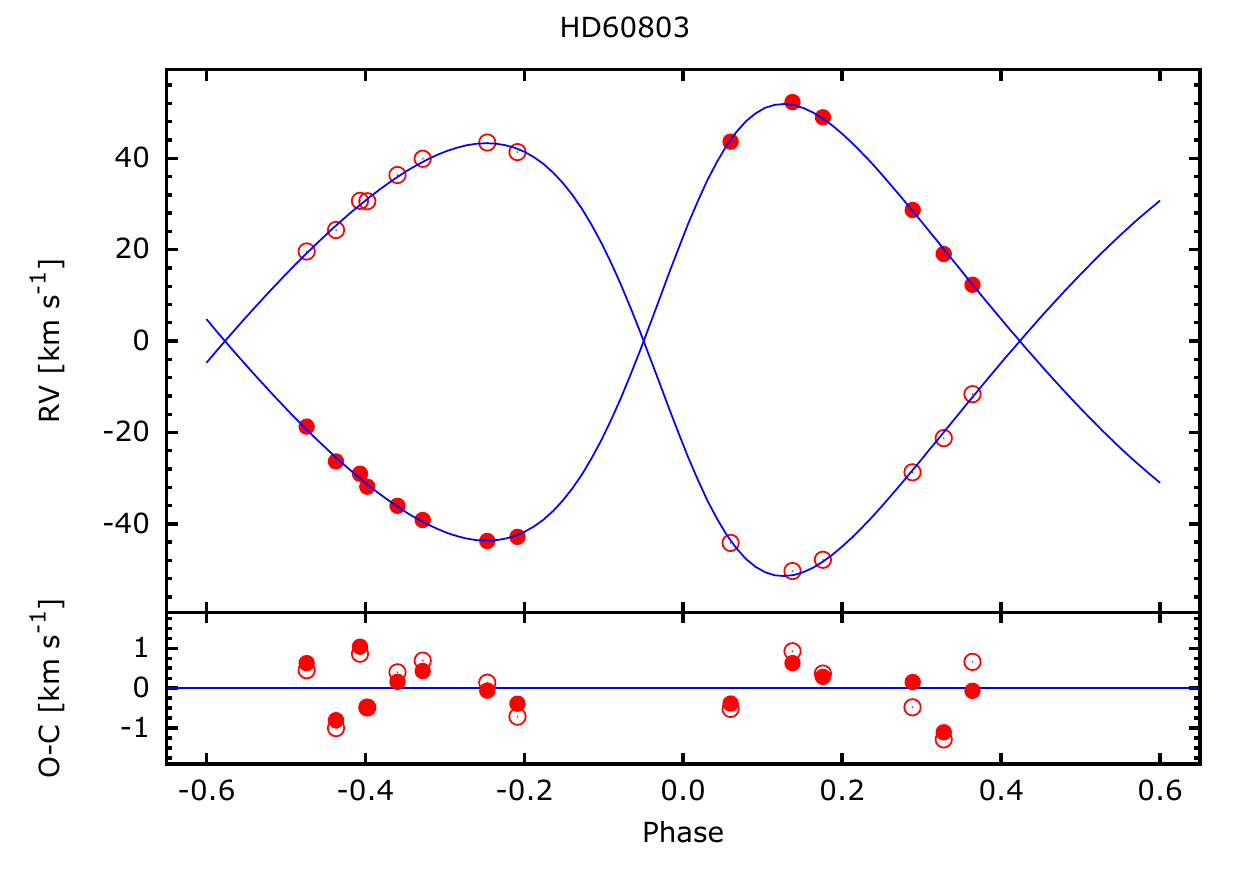}
\includegraphics[width=0.48\textwidth]{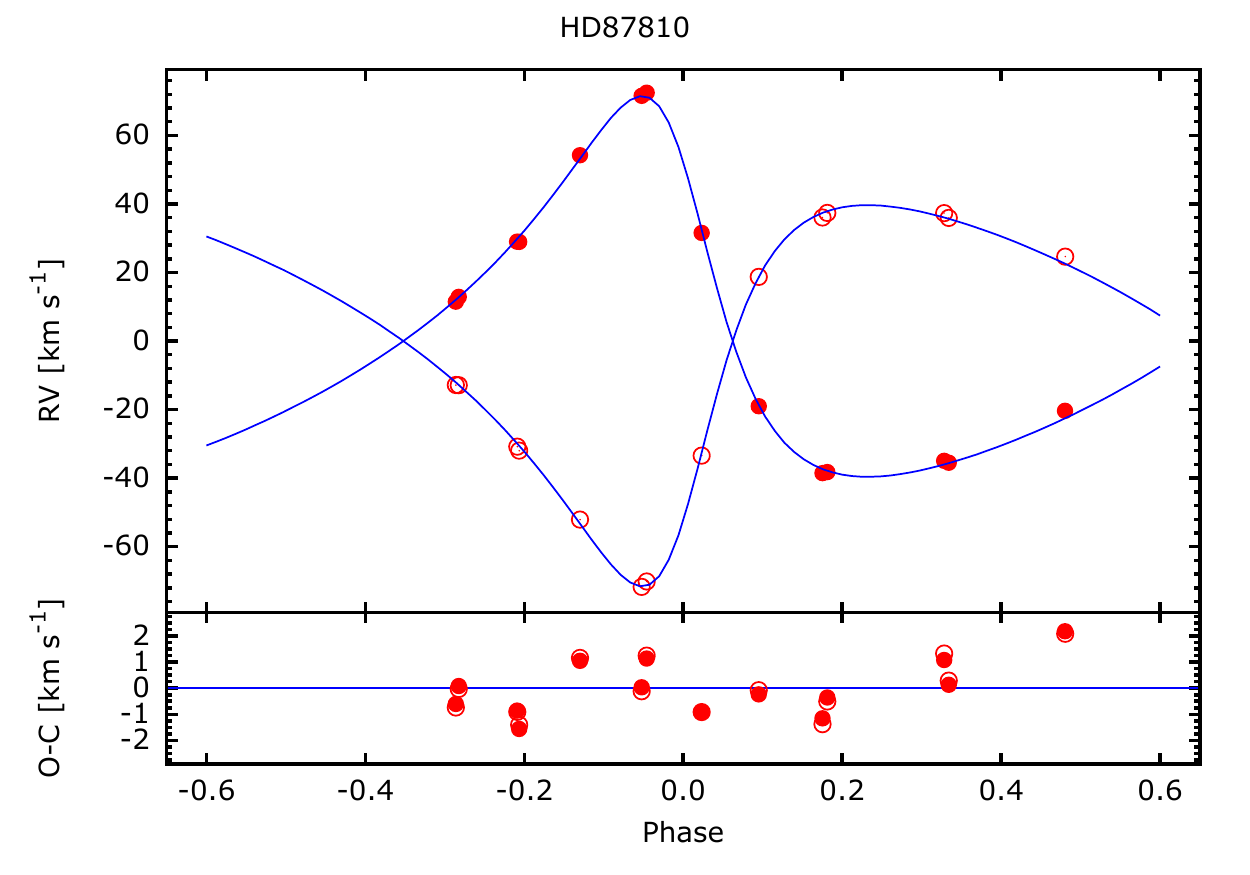}
\caption{RV measurements with the best fit for HD60803 and HD87810. Models (a) from Tab.~\ref{tab:SB2} are shown.}
\label{fig:plotHD60803}
\end{figure*}

Two spectroscopic binaries have been observed during the campaign: HD60803 (G0) and  HD87810 (F3). These targets have been chosen to verify that our approach in terms of data reduction quality and analysis leads to scientifically valuable results. Selection was based on magnitude and coordinate criteria. The obtained orbital solutions have  been compared with those presented in The Ninth Catalogue of Spectroscopic Binary Orbits\footnote{\url{http://sb9.astro.ulb.ac.be/mainform.cgi}} (SB9) by \cite{Pourbaix2004} and summarized in Tab. \ref{tab:SB2}. RV curves with fitted orbital solutions are shown in Fig. \ref{fig:plotHD60803}. Our orbital parameters agree with those presented in the cited publication. Formal errors of period and periastron time in our solutions are significantly higher than what is available in the literature. This is caused by the short time span of our observations. The original RV data for HD60803 has been published by \cite{Griffin1997} and spans 60 years with most data points collected between 1993 and 1996. By keeping the period and periastron time fixed at the values from SB9, we obtain a slightly different set of parameters that are more meaningful in terms of comparing them with the original solution. In both cases, however, the results are consistent. Original RV measurements for HD87810 have been published by \cite{Nordstrom1997} and are based on observations carried out between 1986 and 1991. Following the same approach as for HD60803, we present two sets of results that are consistent with literature data.
\setlength{\tabcolsep}{2pt}
\begin{table*}[htp]
\caption{Orbital solutions for HD60803 and HD87810. Three models are presented: (a) based on this work, (b) based on this work with period and periastron time fixed at literature values and (c) from the SB9 Catalog.}
\begin{center}
\begin{tabular}{crclrclrcl}
\hline
\hline
HD60803			& \multicolumn{3}{c}{This work (a)} & \multicolumn{3}{c}{This work (b)}& \multicolumn{3}{c}{SB9 (c)} \\
\hline
Period (d)					& 	26.20 	&$\pm$& 0.08	& \multicolumn{3}{c}{same as SB9}		&	26.1889	&$\pm$&	0.0006\\
Periastron time (MJD) 		&	49642 	&$\pm$& 23	& \multicolumn{3}{c}{same as SB9}	&	49644.88	&$\pm$&	0.03\\
$e$						&	 0.2177 	&$\pm$& 0.007	& 0.222 &  $\pm$&0.006		&	0.2187	&$\pm$&	0.0017\\
$\omega$	(\arcdeg)			&	113.2 	&$\pm$& 1.6	& 115.794 &  $\pm$& 0.4  		&	113.6	&$\pm$&	0.5\\
$K_1$ (km s$^{-1}$) 			&	47.4		&$\pm$& 0.5	& 47.5 &$\pm$&0.3		&	47.26	&$\pm$&	0.10\\
$K_2$ (km s$^{-1}$) 			&	47.78 	&$\pm$& 0.29	& 47.89&$\pm$&0.29		&	48.16	&$\pm$&	0.12\\
$\gamma$ (km s$^{-1}$) 		&	4.5 		&$\pm$& 0.3	& 4.77&$\pm$&0.25		&	4.60	&$\pm$&	0.06\\
RMS RV$_1$ & \multicolumn{3}{c}{ 0.73} & \multicolumn{3}{c}{ 0.87} &  \multicolumn{3}{c}{ 0.41}\\ 
RMS RV$_2$ & \multicolumn{3}{c}{ 0.59} &  \multicolumn{3}{c}{ 0.54} & \multicolumn{3}{c}{ 0.41}\\ 
RV measurements &  \multicolumn{6}{c}{14}  &  \multicolumn{3}{c}{45} \\
\hline
\end{tabular}
\end{center}
\vspace{0.1cm}
\begin{center}
\begin{tabular}{crclrclrcl}
\hline
\hline
HD87810			& \multicolumn{3}{c}{This work (a)} 	& \multicolumn{3}{c}{This work (b)} 	& \multicolumn{3}{c}{SB9 (c)} \\
\hline
Period (d)			& 	12.948 	&$\pm$& 0.012		&	\multicolumn{3}{c}{same as SB9}	&	12.94724	&$\pm$&	0.00012\\
Periastron time 	(MJD)	&	47441 	&$\pm$& 9	&      \multicolumn{3}{c}{same as SB9}	&	47442.274&$\pm$&	0.008\\
$e$				&	 0.439 	&$\pm$& 0.006		&	0.439	&$\pm$&	0.005	&	0.439	&$\pm$&	0.002\\
$\omega$	(\arcdeg)	&	49.1 		&$\pm$& 1.0		&	49.2	&$\pm$&	0.9	&	48.6		&$\pm$&	0.5\\
$K_1$ (km s$^{-1}$) 	&	55.5		&$\pm$& 0.5		&	55.6	&$\pm$&	0.5		&	55.50	&$\pm$&	0.17\\
$K_2$ (km s$^{-1}$) 	&	55.6	 	&$\pm$& 0.5		&	55.6	&$\pm$&	0.5		&	55.68	&$\pm$&	0.18\\
$\gamma$ (km s$^{-1}$) 	&	-0.5 	&$\pm$& 0.4		&	-0.6	&$\pm$&	0.3		&	-0.24	 	&$\pm$&	0.07\\
RMS RV$_1$ & \multicolumn{3}{c}{1.04} & \multicolumn{3}{c}{1.09} &  \multicolumn{3}{c}{ 0.47}\\ 
RMS RV$_2$ & \multicolumn{3}{c}{1.09} & \multicolumn{3}{c}{1.05} &  \multicolumn{3}{c}{ 0.49}\\ 
RV measurements &  \multicolumn{6}{c}{14}  &  \multicolumn{3}{c}{23} \\
\hline
\end{tabular}
\end{center}
\label{tab:SB2}
\end{table*}
\setlength{\tabcolsep}{6pt}

\subsection{Eclipsing Binaries}

Six detached eclipsing binaries (DEBs) have been observed throughout the campaign. The following sections describe the results of modelling. Derived model parameters are listed in Tab.~\ref{tab:ModelsSolutions}, ASAS light curve and RV plots are shown in Fig. \ref{fig:Plots}.

\subsubsection{ASAS J042724-2756.2, J071626+0548.8, J155259-6637.8} 

These three systems have been previously studied by \cite{Helminiak2009} and observed with the UCLES spectrograph with the
iodine cell on the 3.9-m Anglo-Australian Telescope telescope (AAT) in Siding Spring Observatory, Australia. We compare our results with the previously published models in Tab. \ref{tab:Comparison}. For J155259-6637.8 we present an additional, third model that is based on data we obtained with the HARPS spectrograph on the ESO 3.6-m telescope in La Silla and that has not been previously published. We find that presented models for these three systems are consistent. An extensive model of the J155259-6637.8 binary based on HARPS spectroscopy and our own Solaris photometry will be published in an upcoming paper. The HARPS-based RV curve is shown as the last plot in Fig. \ref{fig:Plots} without the accompanying LC model, since it is identical with the LC model derived for BACHES-based data that is shown above.

J042724-2756.2, also known as CD-28 1554, is a $V=9.89$ mag, F5 eclipsing binary. For this system the errors in mass determination are 3.0\% and 3.5\% for the primary and the secondary components, respectively, compared to  0.94\% and 0.90\% in the cited paper. We have also compared $v_\gamma$ for this system and find that it changed from 7.79(4) to 2.5(5) during 6 years. We have additional data
from different spectrographs that demonstrate that this is a triple system and the change in $v_\gamma$ is due to a third body. With the available photometric data we obtain a model with very small eccentricity and large error of $\omega$. This might suggest that the orbit is circular. If in the analysis we fix $e=0$, the best fit LC model has large residual errors at phases during eclipses. We therefore present a best fit model that assumes a non-circular orbit.

J071626+0548.8, or TYC 176-2950-1 is a G6, $V=10.21$ mag eclipsing binary. It is an eccentric system with $e=0.22$ for which we have derived masses with 3.8\% precision. 

J155259-6637.8, or HD 141344, is a $V=8.99$ mag, F3 type eclipsing binary. Masses for the components are determined with  4.6 and  4.3 \% errors. 

\subsubsection{ASAS J061212-1215.8} 

J061212-1215.8, or HD 42797, is a $V=8.87$ mag, F3 type eclipsing binary that has not been studied in the literature. We have derived masses of the two components with  
4.4\% and 3.6\% errors. With a separation of 12.44 AU, at least the larger component is deformed, what is clearly visible in the presented model. 

\subsubsection{ASAS J072222-1159.8} 

ASAS J072222-1159.8, or HD57797, is a A0 type $V=8.82$ mag eclipsing binary. It has been studied by \cite{Shivvers2014} who present a subset of highly eccentric binaries from the ACVS with $e=0.29\pm0.025$ for this system. This value is consistent with our model ($e=0.289\pm0.018$). Masses have been derived with 3.4\% and 3.3\% errors for the primary and secondary components, respectively. This target has not been observed spectroscopically before, we present its first RV curve and model. 

\subsubsection{ASAS J111134-4956.2} 
J111134-4956.2, or HD97329, is a $V=8.31$ mag A3 type eclipsing binary that has not been studied in the literature before. The two components have masses of 1.92  and 1.70 M$_\Sun$ derived with 10\% errors.

\begin{table*}[ht]
\caption{Comparison of solutions for J042724-2756.2, J071626+0548.8 and J155259-6637.8 obtained with drastically different instrumental setups: BACHES on a 0.5-m telescope, UCLES on a 3.9-m telescope and HARPS on a 3.6-m telescope. Both HARPS and UCLES, compared to BACHES, can be treated as reference instruments.}
\begin{center}
\small
\begin{tabular}{lrrrrrrrrrrrr}
\hline
\hline
~ & $T_0$ & $P$ & $K_1$ & $K_2$ & $M_1$ & $M_2$  & $e$ & $\omega$ & $i$\\ 
~ & (JD)    &  (d)  &  (km s$^{-1}$) &  (km s$^{-1}$) & (M$_\Sun$) & (M$_\Sun$) & --- & (\arcdeg) & (\arcdeg)  \\ \hline	
\multicolumn{10}{l}{J042724-2756.2}																			\\	
BACHES	&	2451882.646	&	8.94657	&	74.0	&	69.2	&	1.33	&	1.42	&	0.011	&	173	&	85.6	\\	
~	&	0.006	&	0.00004	&	1.2	&	0.7	&	0.04	&	0.05			&	0.016	&	80		&	2.8	\\	
UCLES	&	2451882.653 	&	8.94657 	&	74.0	&	70.7	&	1.383 	&	1.449 	&	0.012	&	238	&	86	\\	
~	&	0.007	&	0.00005	&	0.4	&	0.4	&	0.013	&	0.013	&	0.04	&	24	&	1	\\	\hline
\multicolumn{10}{l}{J071626+0548.8}																			\\	
BACHES	&	 2452426.010	&	11.526	&	 62	&	62.1	&	1.06	&	1.06	&	0.21	&	49.3	&	88.2	\\	
~	&	0.004	&	0.023	&	1	&	1.1	&	0.04	&	0.04	&	0.01	&	2.1	&	1.8	\\	
UCLES	&	2452426.025 	&	11.55478	&	61.2	&	62.1	&	1.055	&	1.041	&	0.21	&	52	&	88.2	\\	
~	&	0.008	&	0.00006	&	0.3	&	0.3	&	0.008	&	0.008	&	0.05	&	10	&	0.7	\\	\hline
\multicolumn{10}{l}{J155259-6637.8}																			\\	
BACHES	&	2451934.524	&	5.744756	&	93.9	&	76.0	&	1.31	&	1.62	&	0.0	&	---	&	87.5	\\	
~	&	0.007	&	0.000023	&	1.8	&	1.6	&	0.06	&	0.07	&	(fixed)	&	---	&	1.2	\\	
UCLES	&	2451934.536	&	5.744754 	&	93.47	&	76.98	&	1.352	&	1.644	&	0.0	&	---	&	84.5 	\\	
~	&	0.006	&	0.000022	&	0.12	&	0.06	&	0.003	&	0.004	&	(fixed)	&	---	&	0.08	\\	
HARPS	&	2451934.536	&	5.744764	&	93.55	&	77.105	&	1.342	&	1.628	&	0.0	&	---	&	87.08	\\	
~	&	0.007	&	0.000007	&	0.04	&	0.026	&	0.004	&	0.005	&	(fixed)	&	---	&	1.2	\\	\hline
\hline
\end{tabular}
\end{center}
\label{tab:Comparison}
\end{table*}

\begin{table*}[ht]
\caption{Solutions for eclipsing binaries. Formal errors are noted directly under the parameter value. For effective temperatures ($T_1$ and $T_2$) only the error for $T_2$ is given, since $T_1$ is fixed in the fitting process.}
\begin{center}
\tiny{
\begin{tabular}{rlrrrrrr}
\hline
\hline
	Parameter	&	Unit	&	J042724-2756.2	&	J061212-1215.8	&	J071626+0548.8	&	J072222-1159.8	&	J111134-4956.2	&	J155259-6637.8	\\	\hline	
	$T_0$	&	(JD)	&	2451882.646	&	2451869.095	&	 2452426.010	&	2451890.526	&	2451873.963	&	2451934.524	\\ 		 \vspace{0.1cm}
		&		&	0.006	&	0.006	&	0.004	&	0.015	&	0.010	&	0.007	\\		
	$P$	&	(d)	&	8.94657	&	2.945024	&	11.526	&	13.0147	&	7.73576	&	5.744756	\\		 \vspace{0.1cm}
		&		&	0.00004	&	0.000019	&	0.023	&	0.00010	&	0.00004	&	0.000023	\\		
	$K_1$	&	(km s$^{-1}$) 	&	74.0	&	113.6	&	 62	&	69.7	&	78	&	93.9	\\		 \vspace{0.1cm}
		&		&	1.2	&	1.0	&	1	&	0.8	&	4	&	1.8	\\		
	$K_2$	&	(km s$^{-1}$) 	&	69.2	&	96.0	&	62.1	&	79.0	&	88	&	76.0	\\		 \vspace{0.1cm}
		&		&	0.7	&	1.1	&	1.1	&	1.1	&	4	&	1.6	\\		
	$e$	&		&	0.009	&	0.0	&	0.21	&	0.289	&	0.150	&	0.0	\\		 \vspace{0.1cm}
		&		&	0.092	&	(fixed)	&	0.01	&	0.018	&	0.002	&	(fixed)	\\		
	$i$	&	(deg.)	&	85.6	&	79	&	88.2	&	87	&	83.1	&	87.5	\\		 \vspace{0.1cm}
		&		&	2.8	&	4	&	1.8	&	2	&	0.7	&	1.2	\\		
	$a$	&	(R$_\sun$)	&	25.39	&	12.44	&	27.6	&	36.7	&	25.3	&	19.30	\\		 \vspace{0.1cm}
		&		&	0.26	&	0.19	&	0.3	&	0.4	&	0.8	&	0.28	\\		
	$\omega$	&	(\arcdeg)	&	173	&	---	&	49.3	&	14.6	&	82	&	---	\\		 \vspace{0.1cm}
		&		&	80	&	---	&	2.1	&	1.7	&	4	&	---	\\		
	$v_\gamma$	&	(km s$^{-1}$) 	&	2.6	&	-15.4	&	12.6	&	19.7	&	7.2	&	10.5	\\		 \vspace{0.1cm}
		&		&	0.5	&	 0.9	&	0.5	&	0.5	&	1.4	&	0.8	\\		
	RMS RV1	&	(km s$^{-1}$) 	&	2.69	&	3.27	&	1.34	&	1.45	&	2.58	&	1.79	\\ 		 \vspace{0.1cm}
	RMS RV2	&	(km s$^{-1}$) 	&	4.38	&	3.29	&	1.62	&	2.82	&	1.60	&	1.58	\\		
	$T_1$	&	(K)	&	6175	&	6263	&	6034	&	7057	&	6855	&	5975	\\		
	$T_2$	&	(K)	&	6024	&	6034	&	5966	&	6455	&	6021	&	6242	\\		 \vspace{0.1cm}
	~	&		&	40	&	60	&	100	&	40	&	30	&	20	\\		
	$M_1$	&	(M$_\sun$)	&	1.33	&	1.37	&	1.06	&	2.08	&	1.92	&	1.31	\\		 \vspace{0.1cm}
		&		&	0.04	&	0.06	&	0.04	&	0.07	&	0.20	&	0.06	\\		
	$M_2$	&	(M$_\sun$)	&	1.42	&	1.63	&	1.06	&	1.83	&	1.70	&	1.62	\\		 \vspace{0.1cm}
		&		&	0.05	&	0.07	&	0.04	&	0.06	&	0.17	&	0.07	\\		
	$R_1$	&	(R$_\sun$)	&	1.5	&	1.9	&	1.18	&	3.4	&	3.5	&	1.5	\\		 \vspace{0.1cm}
		&		&	0.8	&	0.8	&	0.05	&	0.4	&	0.4	&	0.11	\\		
	$R_2$	&	(R$_\sun$)	&	2.5	&	2.5	&	1.67	&	1.8	&	1.89	&	2.72	\\		 \vspace{0.1cm}
		&		&	0.9	&	0.5	&	0.13	&	0.6	&	0.12	&	0.13	\\		
	log $g_1$	&	(cm s$^{-1}$)	&	4.2	&	4.0	&	4.3	&	3.68	&	3.64	&	4.19	\\		 \vspace{0.1cm}
		&		&	0.5	&	0.4	&	0.03	&	0.11	&	0.10	&	0.06	\\		
	log $g_2$	&	(cm s$^{-1}$)	&	3.8	&	3.9	&	4.02	&	4.2	&	4.12	&	3.78	\\		 \vspace{0.1cm}
		&		&	0.3	&	0.16	&	0.07	&	0.3	&	0.05	&	0.04	\\		
	$v_{synchr, 1}$	&	(km s$^{-1}$) 	&	9	&	32	&	5.17	&	13.3	&	22.7	&	13.4	\\		 \vspace{0.1cm}
		&		&	4	&	13	&	0.23	&	1.7	&	2.7	&	1.0	\\		
	$v_{synchr, 2}$	&	(km s$^{-1}$) 	&	14	&	43	&	7.3	&	6.9	&	12.3	&	24.0	\\		 \vspace{0.1cm}
		&		&	5	&	7	&	0.6	&	2.4	&	0.8	&	1.2	\\		
	log $L_1$	&	(L$_\sun$)	&	0.5	&	0.7	&	0.22	&	1.42	&	1.38	&	0.43	\\		 \vspace{0.1cm}
		&		&	0.5	&	0.4	&	0.06	&	0.12	&	0.12	&	0.08	\\		
	log $L_2$	&	(L$_\sun$)	&	0.9	&	0.87	&	0.50	&	0.7	&	0.63	&	1.01	\\		 \vspace{0.1cm}
		&		&	0.3	&	0.16	&	0.08	&	0.3	&	0.08	&	0.06	\\		
	$M_{bol,1}$	&	(mag)	&	3.5	&	3.0	&	4.20	&	1.20	&	1.30	&	3.68	\\		 \vspace{0.1cm}
		&		&	1.3	&	0.9	&	0.14	&	0.30	&	0.29	&	0.20	\\		
	$M_{bol,2}$	&	(mag)	&	2.6	&	2.6	&	3.50	&	3.0	&	3.18	&	2.23	\\		 \vspace{0.1cm}
		&		&	0.8	&	0.4	&	0.20	&	0.8	&	0.19	&	0.15	\\					
		
\end{tabular}
	}
\end{center}
\label{tab:ModelsSolutions}
\end{table*}

\begin{figure*}[ht]
\begin{center}
\includegraphics[width=0.48\textwidth]{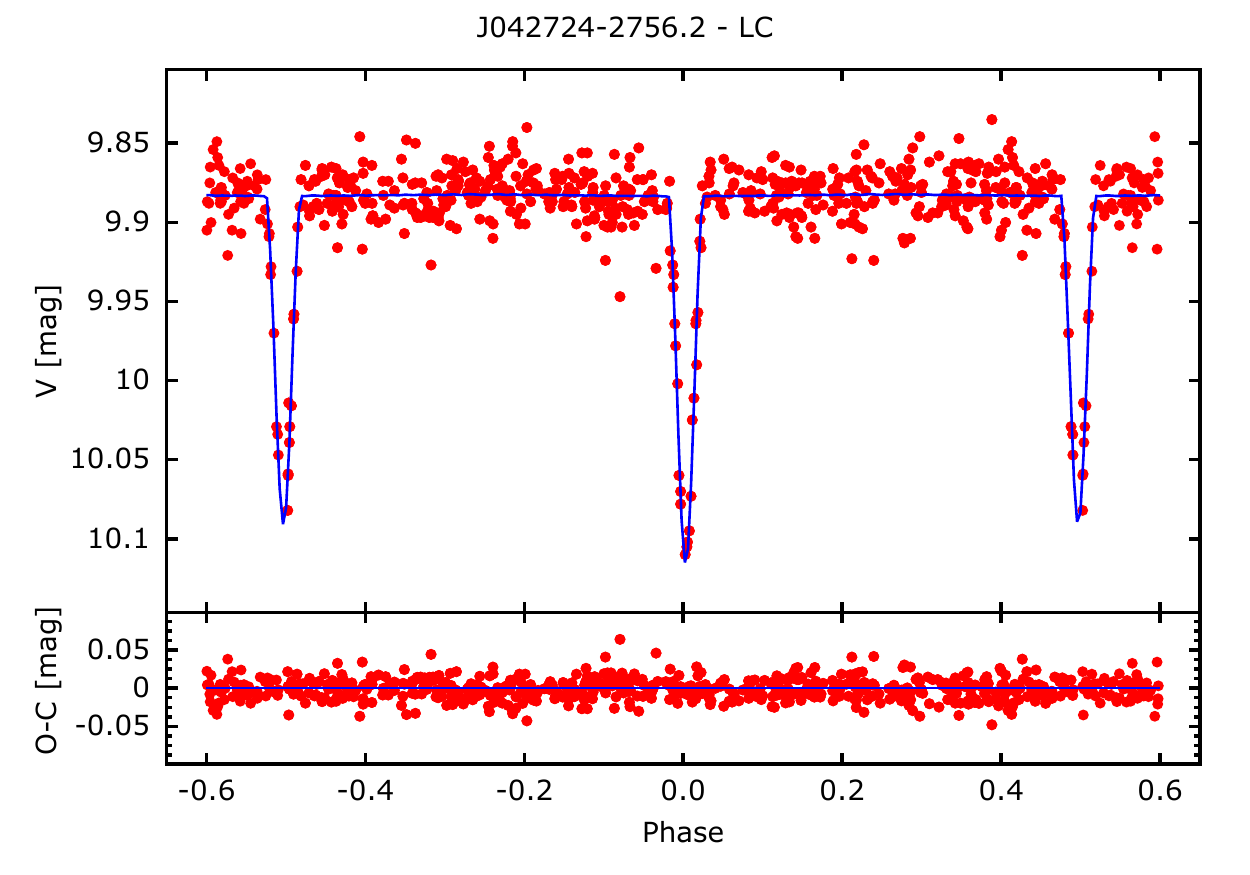}
\includegraphics[width=0.48\textwidth]{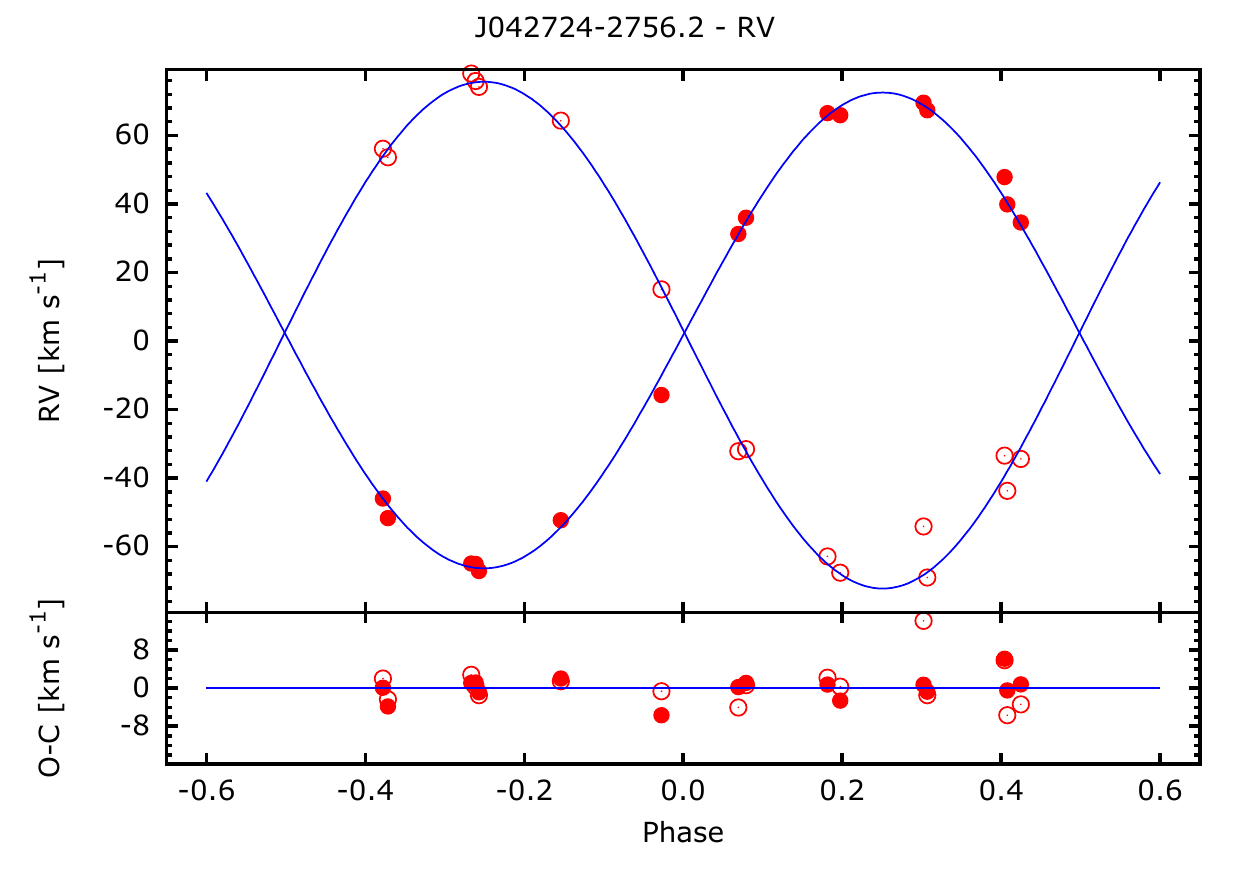}
\includegraphics[width=0.48\textwidth]{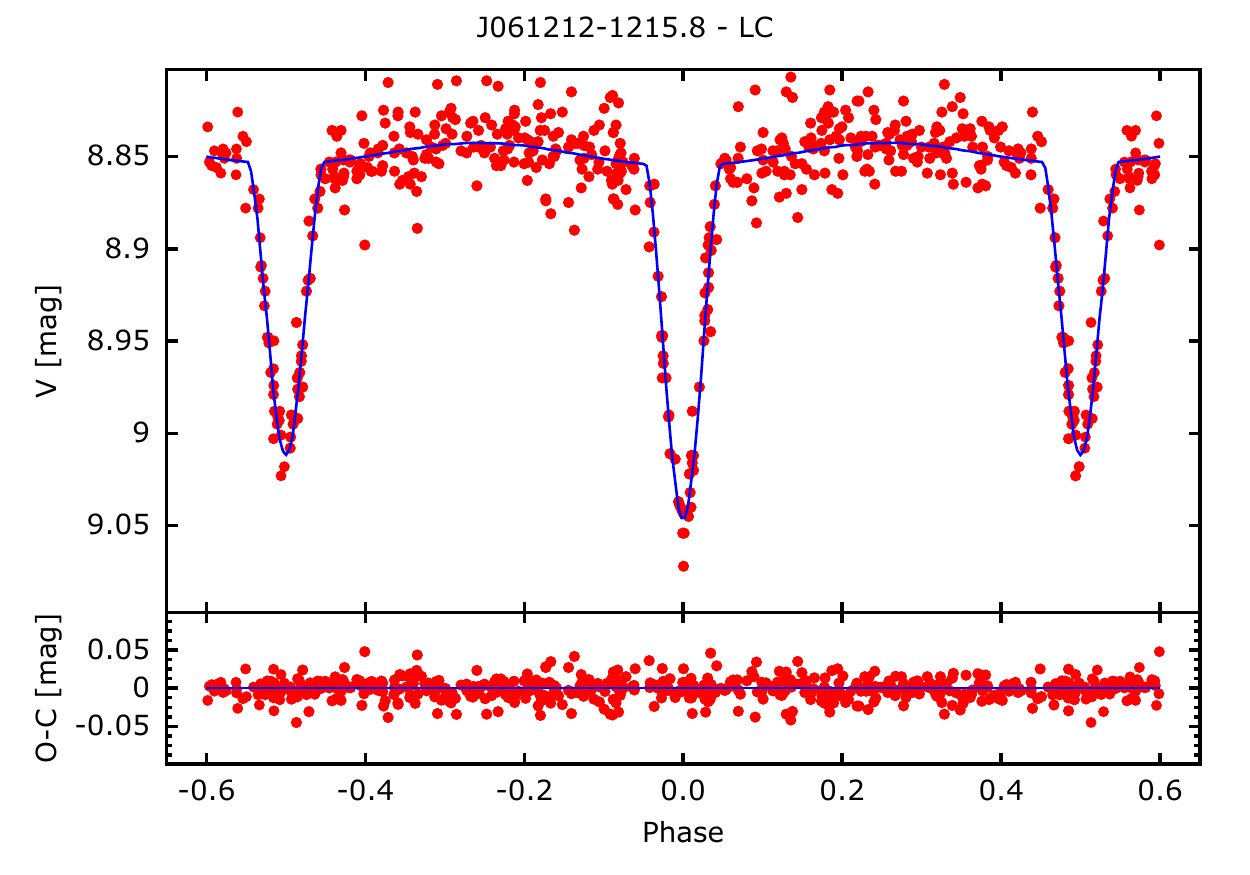}
\includegraphics[width=0.48\textwidth]{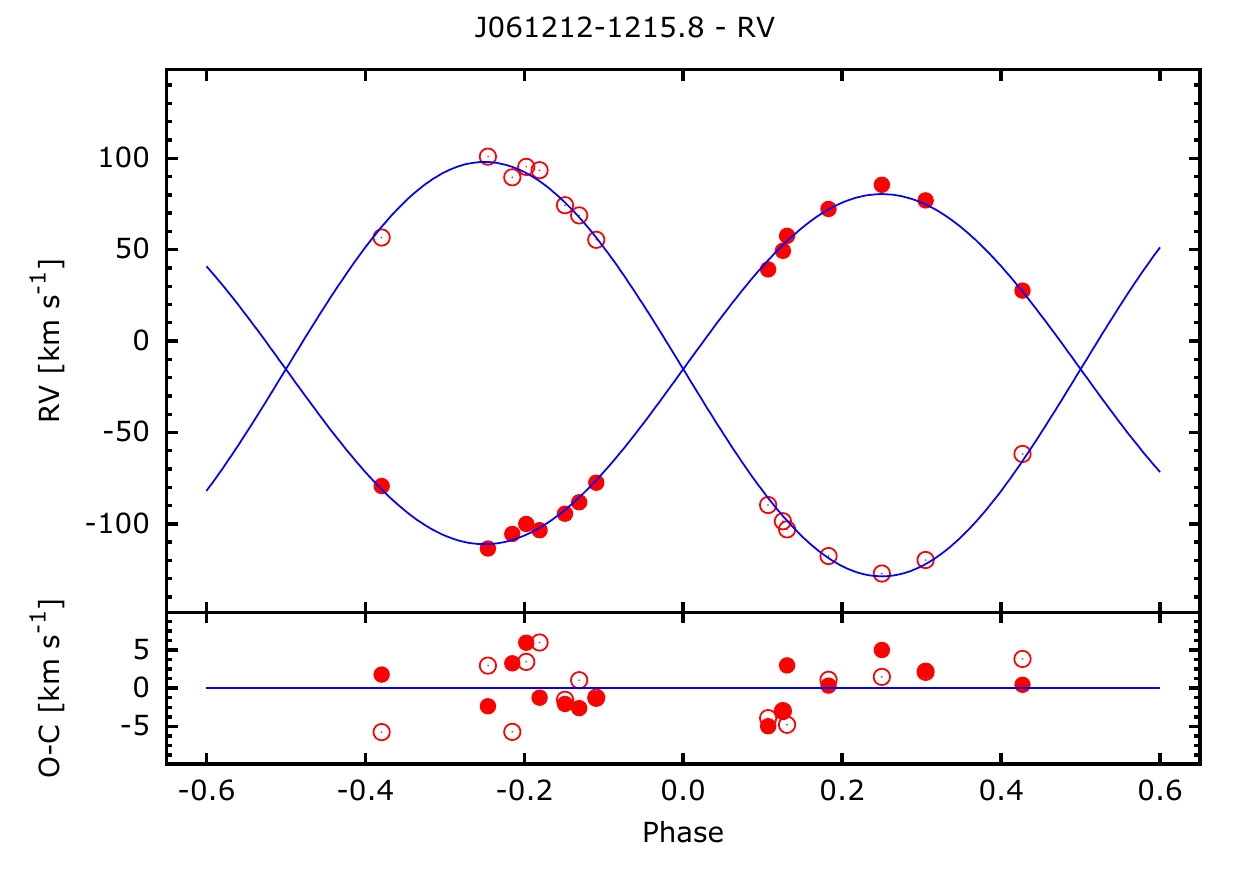}
\includegraphics[width=0.48\textwidth]{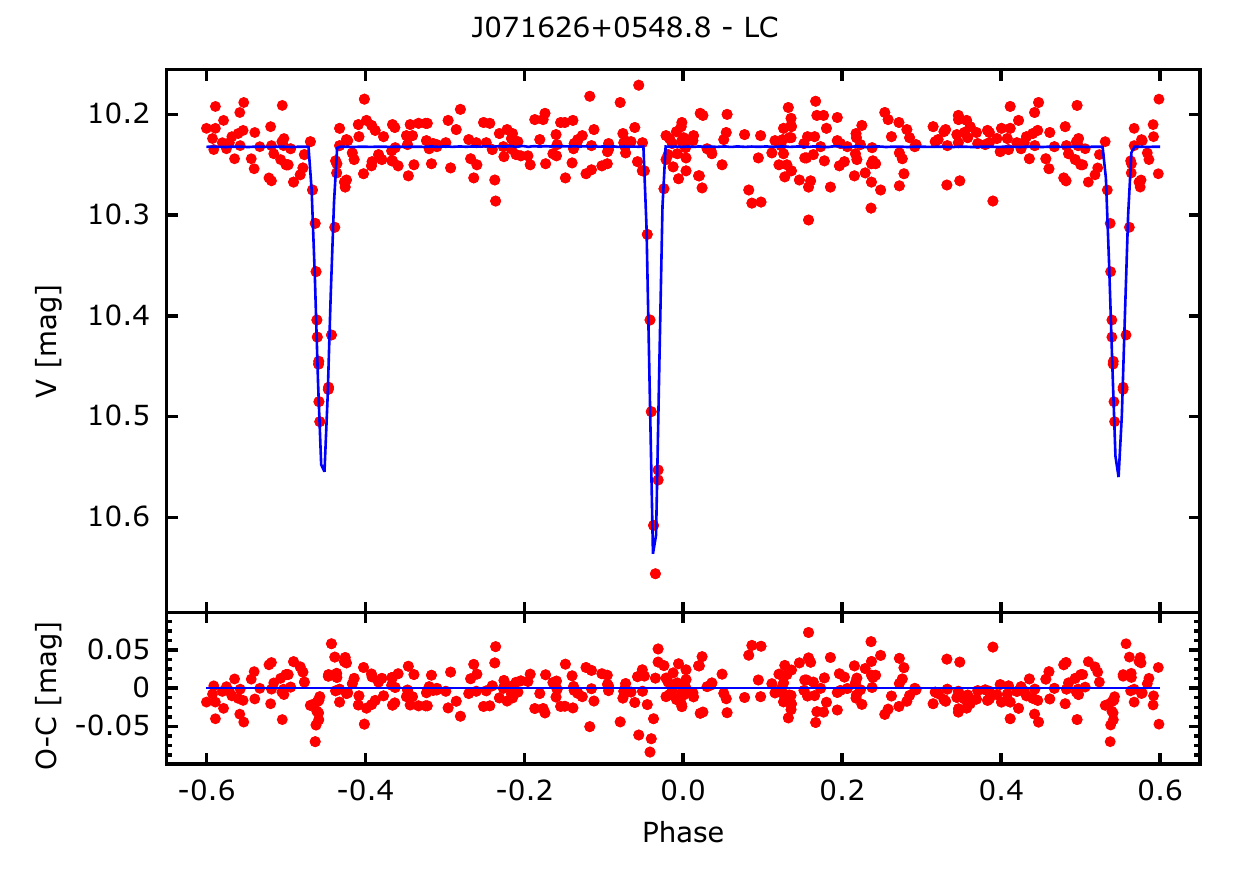}
\includegraphics[width=0.48\textwidth]{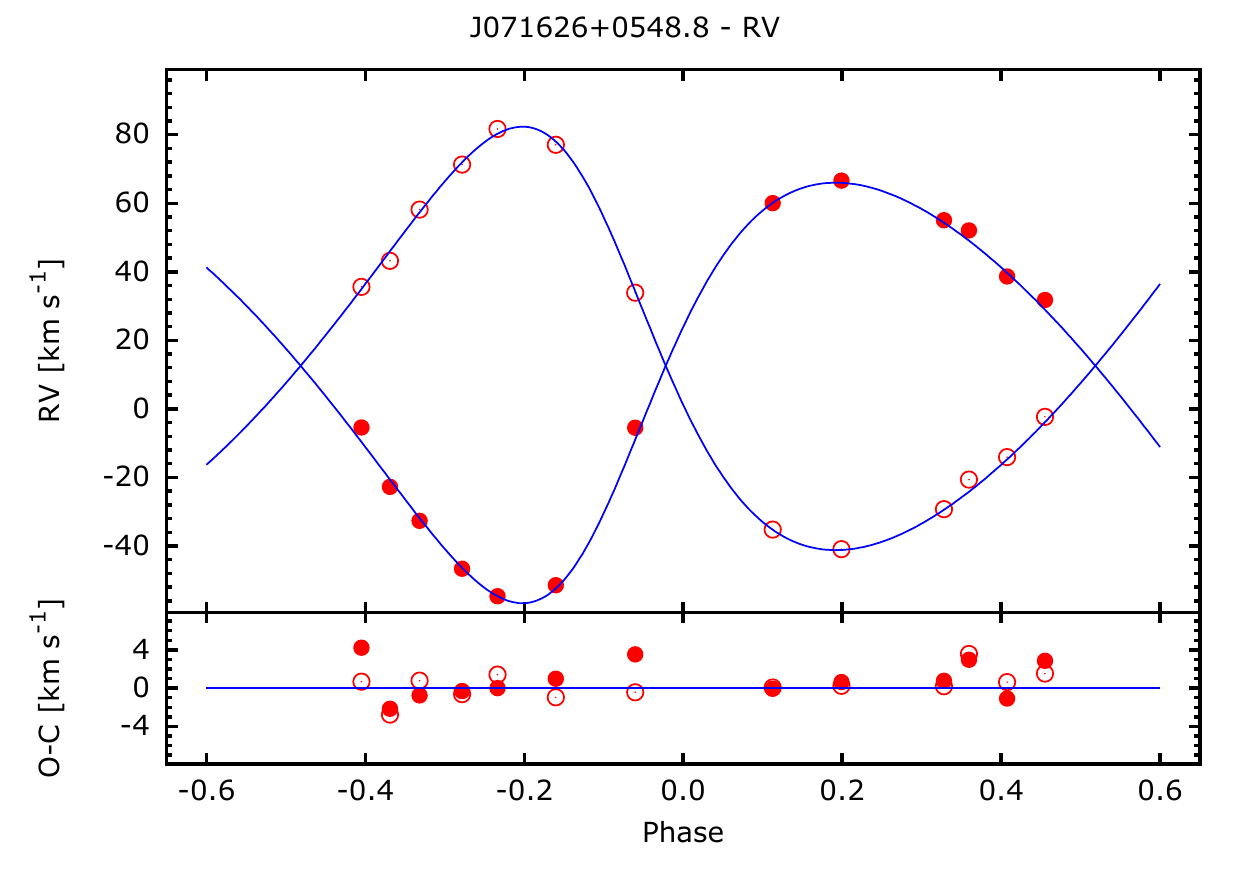}
\end{center}
\caption{Light curves and RV plots for eclipsing binaries along with derived models.}
\label{fig:Plots}  
\end{figure*}
\addtocounter{figure}{-1}
\begin{figure*}[ht]
\begin{center}
\includegraphics[width=0.45\textwidth]{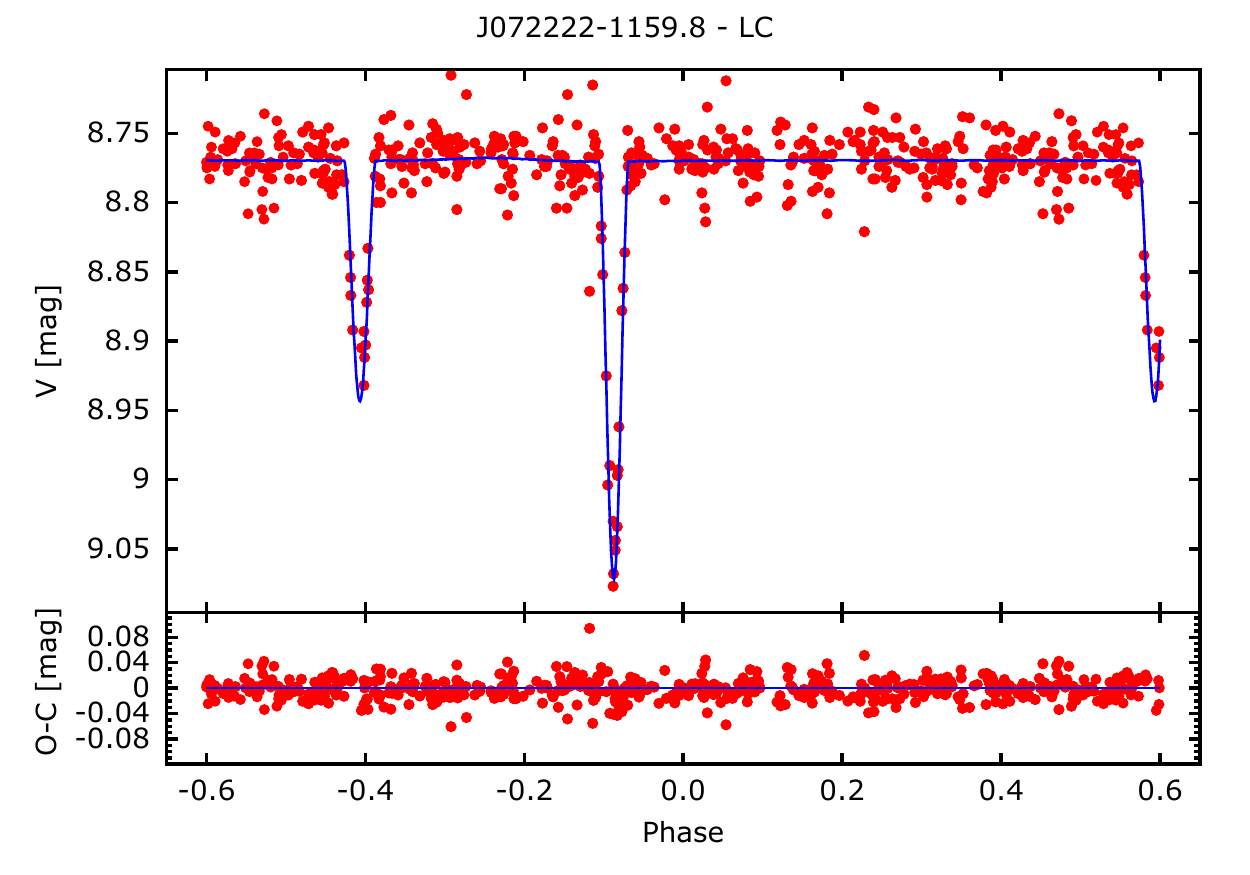}
\includegraphics[width=0.45\textwidth]{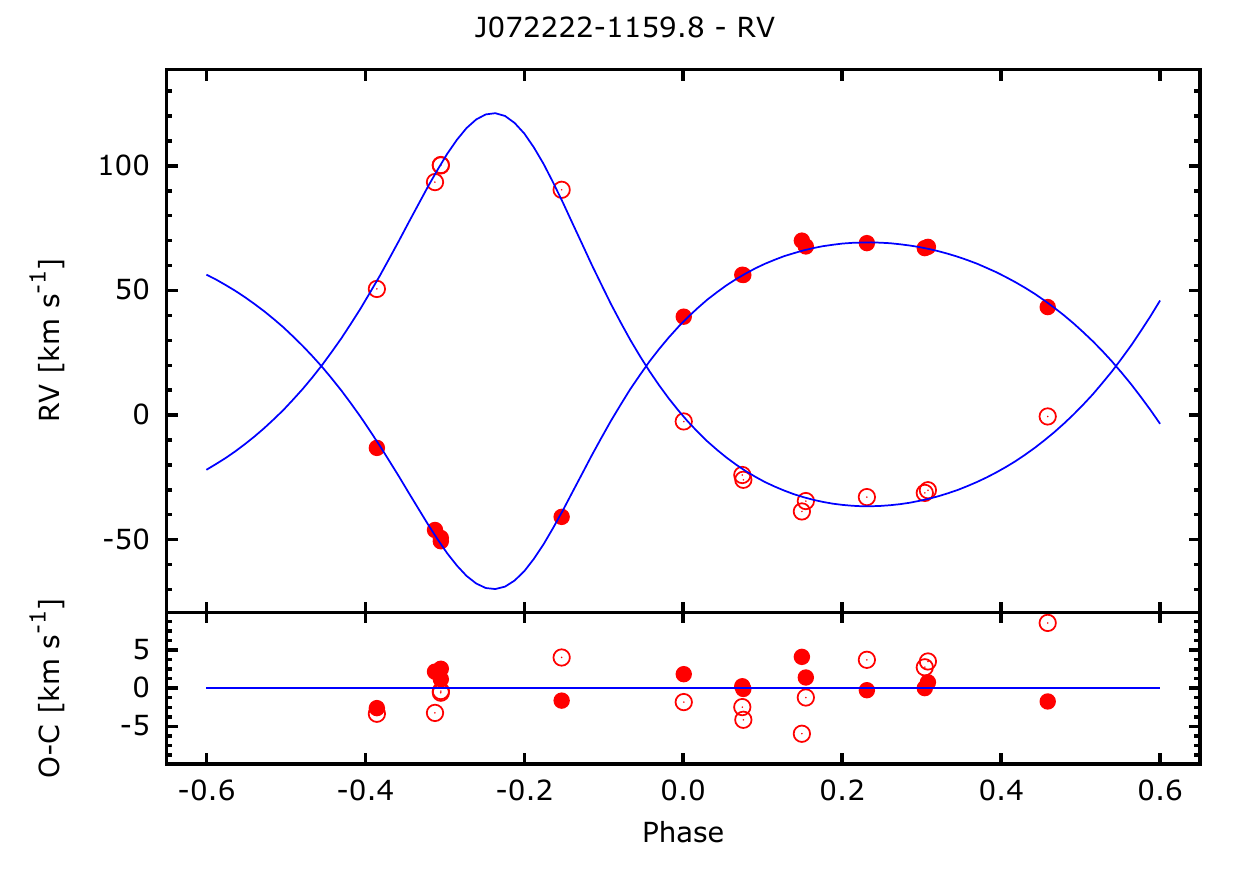}
\includegraphics[width=0.45\textwidth]{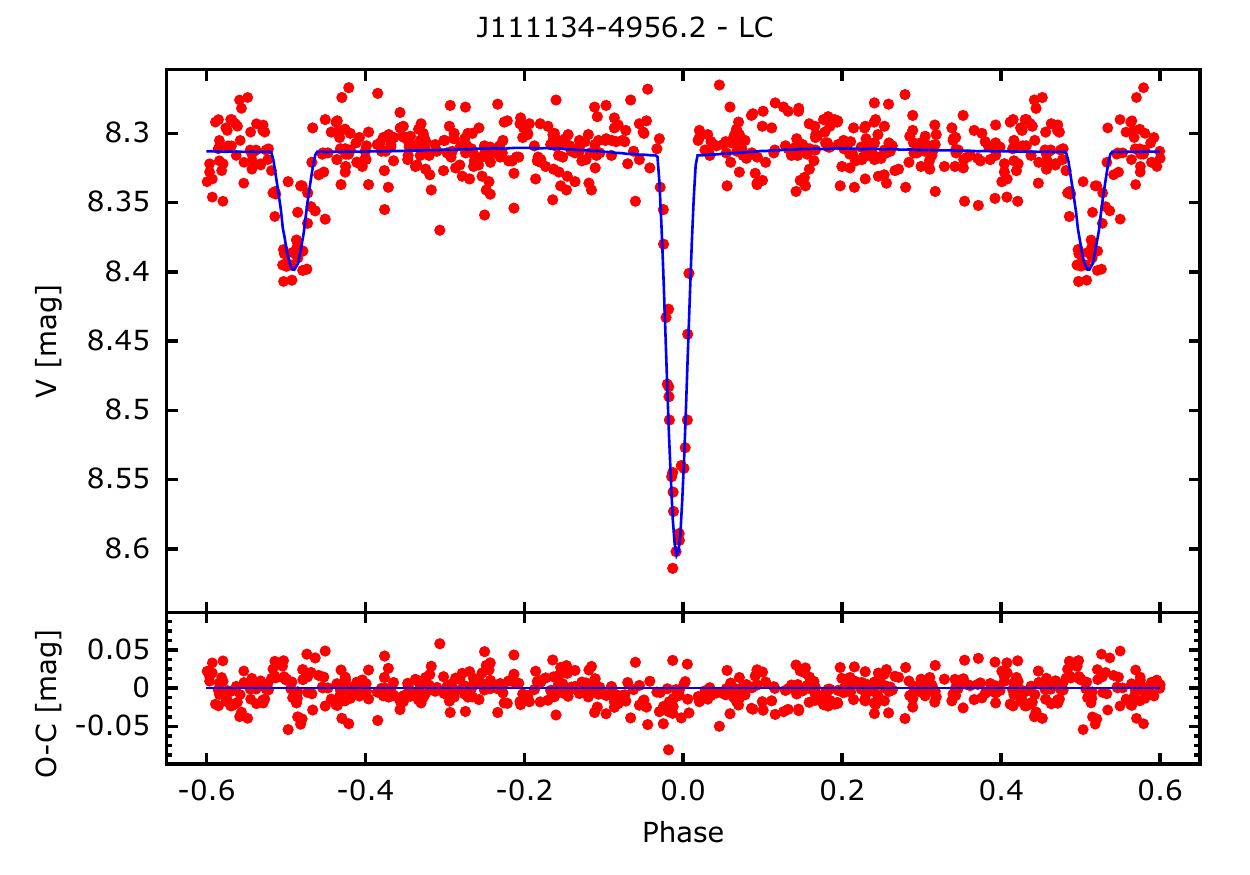}
\includegraphics[width=0.45\textwidth]{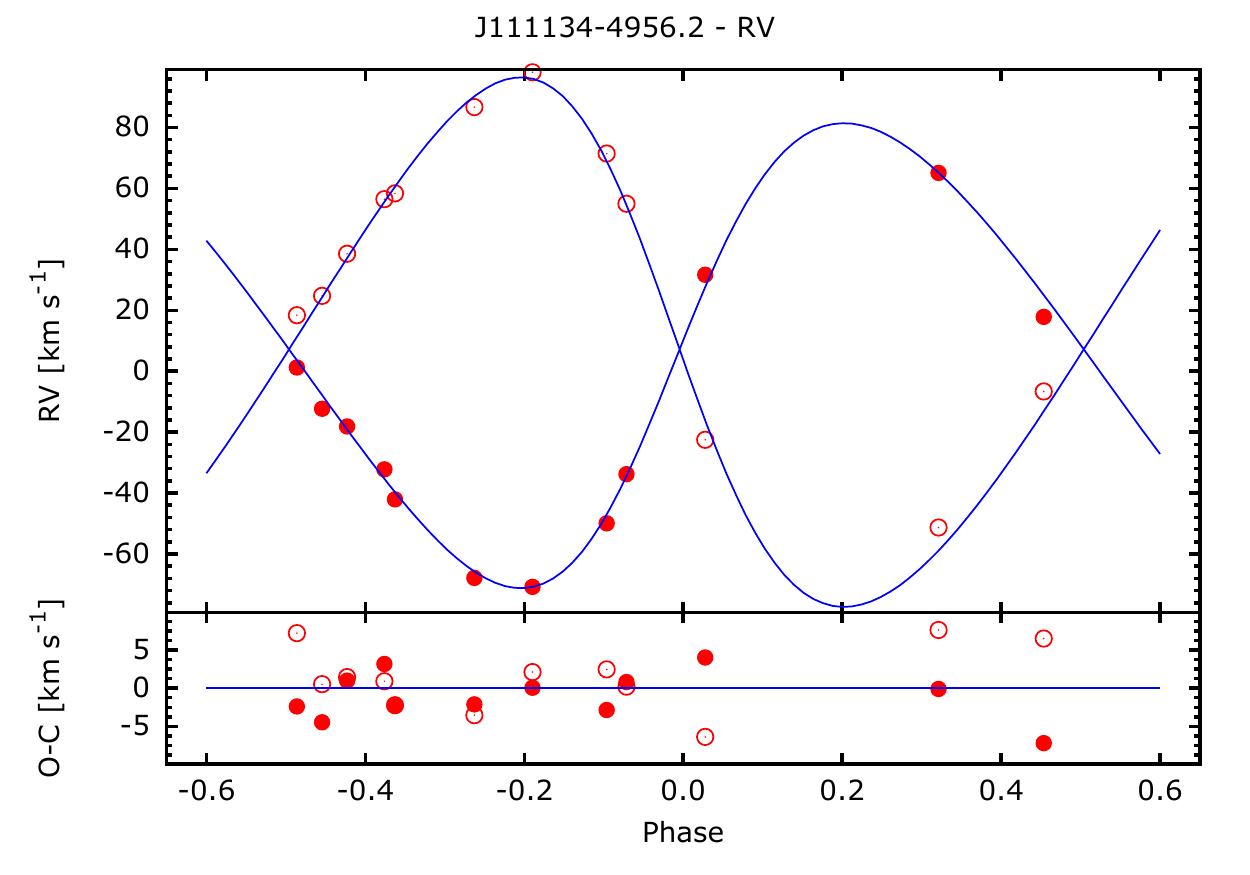}
\includegraphics[width=0.45\textwidth]{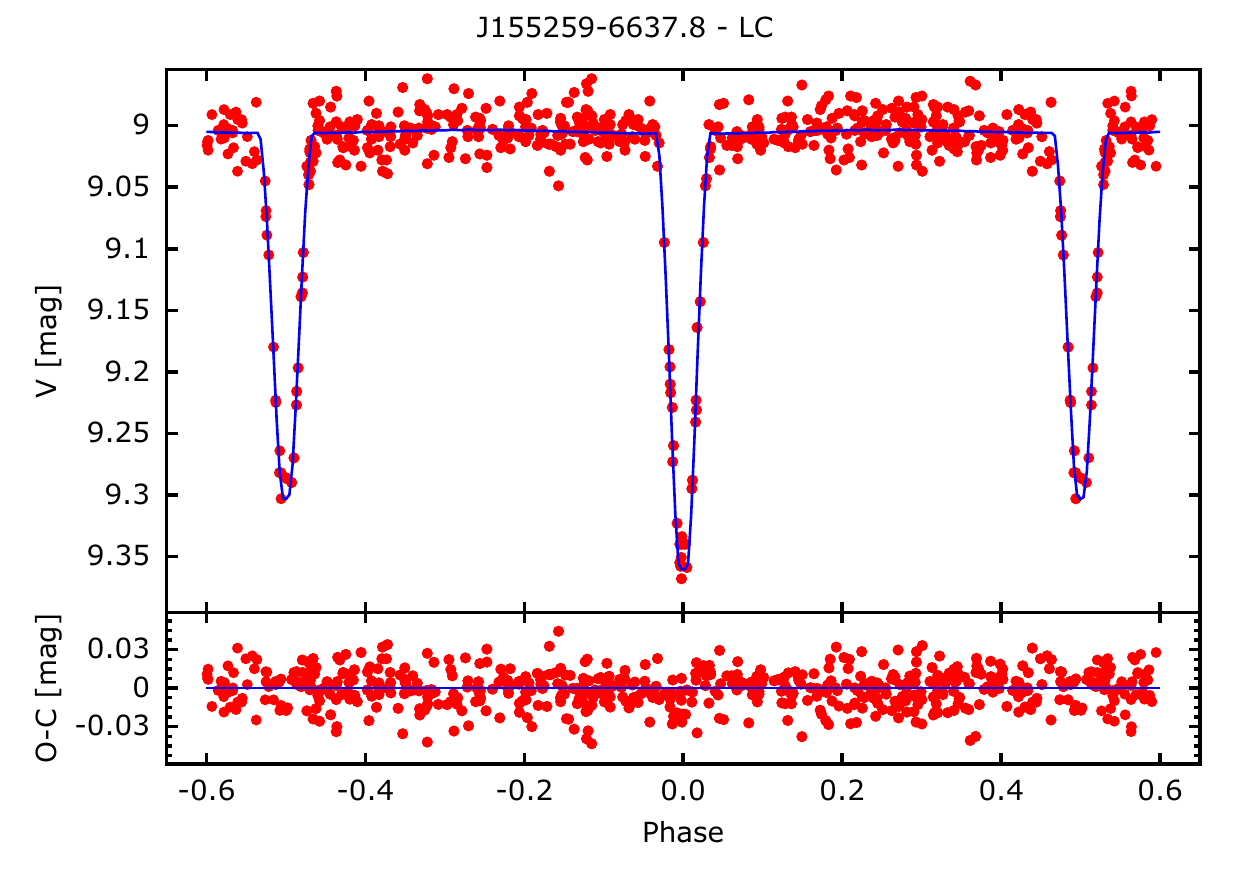}
\includegraphics[width=0.45\textwidth]{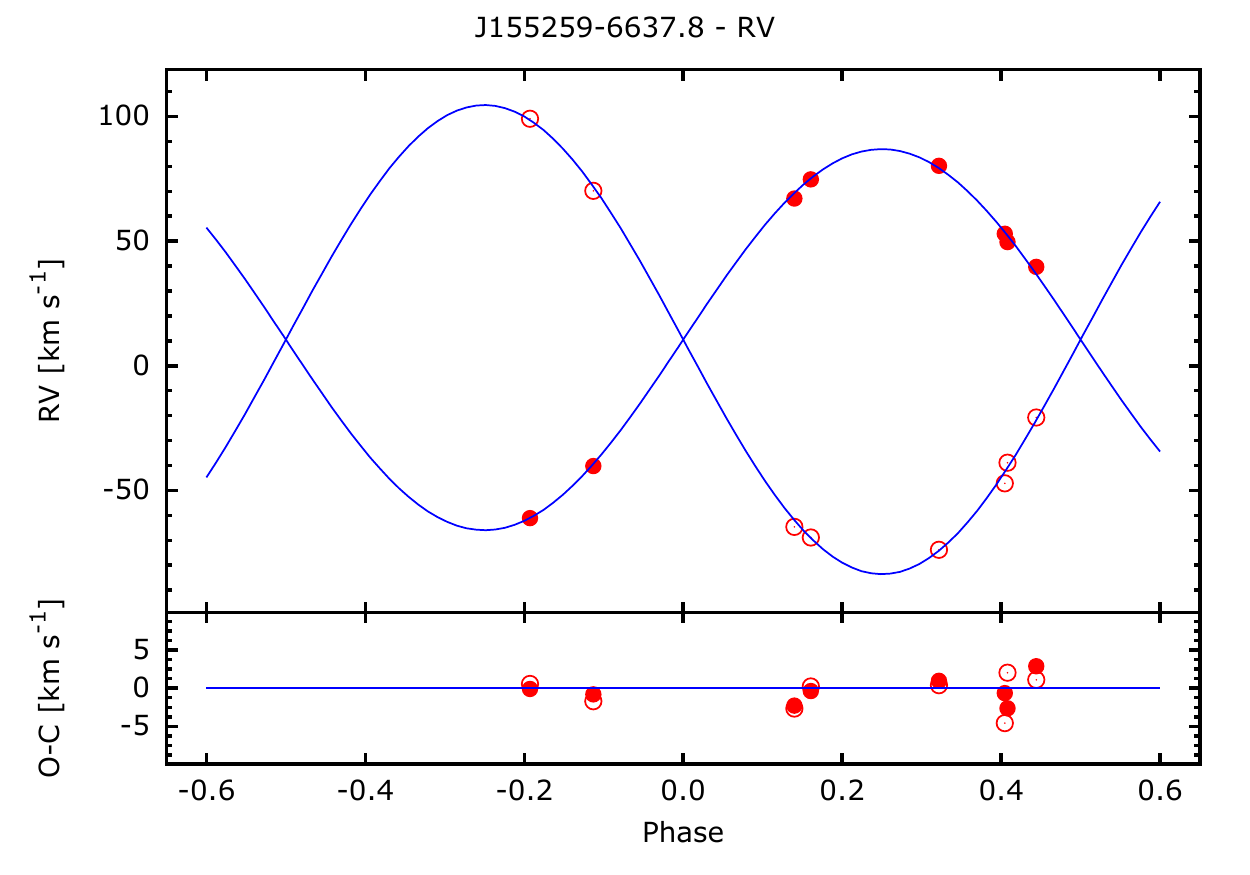}
\includegraphics[width=0.45\textwidth]{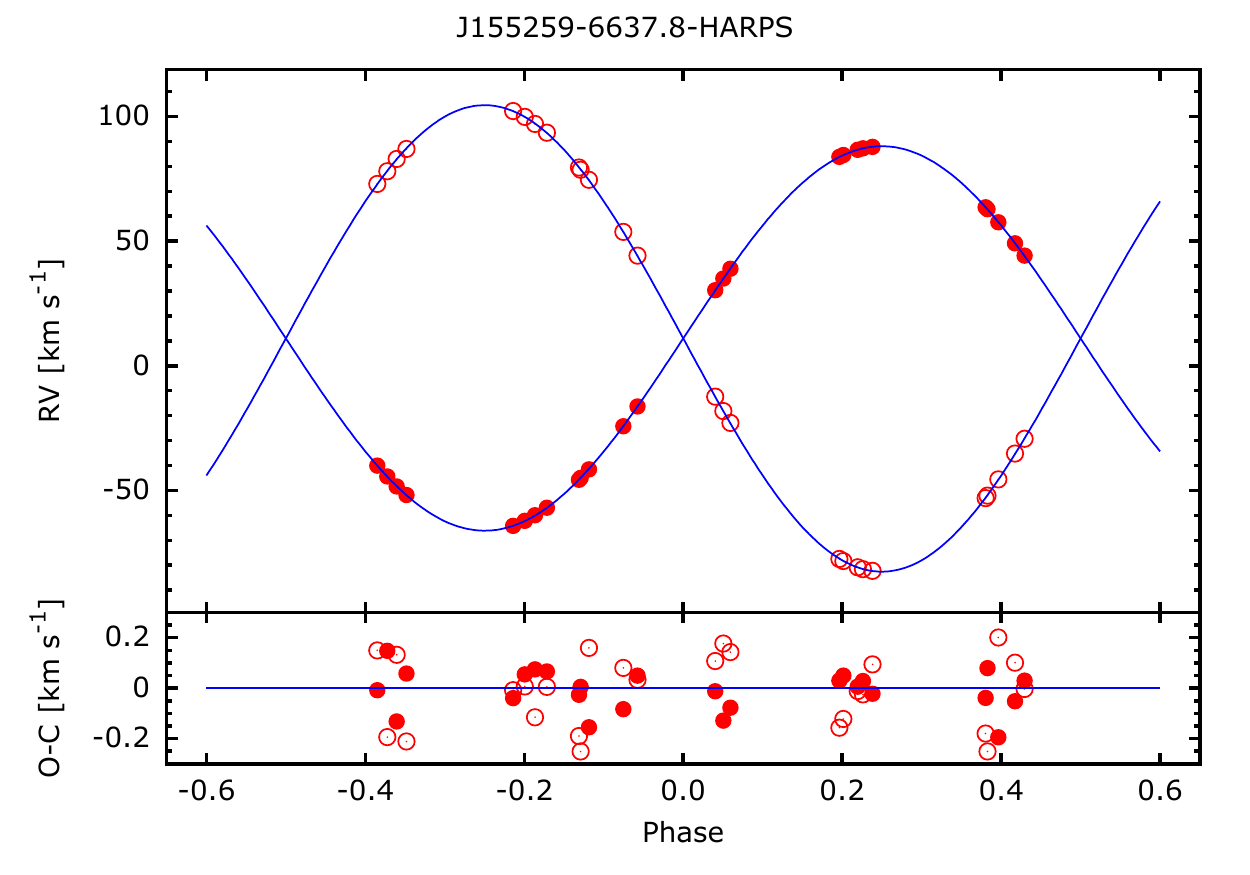}
\end{center}
\caption{(continued)}
\label{fig:cont}
\end{figure*} 

%
%
%
%
%

\section{Summary and Discussion}
\label{sec:Summary}
We have presented results of the scientific commissioning of a low-cost \'{e}chelle spectrograph. The main purpose of this instrument is to conduct a spectroscopic survey of eclipsing binaries. We have shown that a small slit spectrograph attached directly to a 0.5-m telescope can be a very capable instrument. The introductory campaign covered ten targets that were observed from February until April 2015 in a semiautomatic manner. A total of 146 spectra have been acquired for two spectroscopic standards, two spectroscopic binaries and six eclipsing binaries. The targets have been chosen in such a way to test the capabilities of the instrument, its stability, and efficiency. We were able to optimize the data acquisition process and refine the data reduction procedures. As a result we have derived models for the six DEBs, three of which have not been studied spectroscopically in the literature before. Obtained stellar parameters of the remaining four have been compared with literature data. We find that in a on average we can achieve an SNR of at least 22 for the brightest part of the spectrum for a $V=10$ mag DEB during a 30 min. exposure. Having a set of $\sim$10 similar quality spectra yields an orbital solution that gives masses of the components with $\sim$3\% errors for specific cases. As pointed out by \cite{Blake2008}, this is sufficient to test stellar evolution codes. Although the instrument's stability suffers from ambient temperature changes that correspond to an amplitude of $\sim25$ km s$^{-1}$ throughout the campaign, ThAr calibration spectra allow us to achieve orbital fit RMS better than 1.5 km s$^{-1}$ for some targets. This is consistent with results obtained for spectroscopic standards. 

As noted in Sec. 5.2., errors of radial velocity measurements are increased in quadrature to obtain an orbital fit with reduced $\chi^2$ $\approx$ 1. This approach is conservative and guarantees that the errors will not be underestimated. Photometric errors that are associated with the ACVS photometric data are used by both  in \textsc{jktebop} and \textsc{phoebe} codes. Once a model is solved, i.e. values of parameters obtained with \textsc{v2fit}, \textsc{jktebop} and \textsc{phoebe} are consistent (\textsc{v2fit} deals with the RV only, \textsc{jktebop} with LC only and \textsc{phoebe} combines both), which often requires several iterations, the final values are used in \textsc{jktabsdim} to compute absolute values of parameters of the binary system. 
This formal approach can be easily defended when comparing our results with those obtained with different instrumental setups. In this paper we have demonstrated that models based on our spectroscopic data agree in terms of absolute parameters' values with solutions based on radial velocity measurements obtained from much more advanced instruments. Specifically, we have included a model for the eclipsing binary J155259-6637.8 that is based on ACVS photometry and RV data from HARPS. RV precision is the deciding factor in terms of mass uncertainties Therefore, a model derived from very precise RV measurements, such as those from HARPS, can be treated as an absolute benchmark. The computed masses are 1.342 M$_\Sun$ and 1.628M$_\Sun$ with $\sim$0.3\% errors. For BACHES data we obtain 1.31 M$_\Sun$  and 1.62 M$_\Sun$  with $\sim$4.5\% formal errors. Comparing these with benchmark values we see that they differ by 2.4\% and 0.5\%, i.e. well within formal errors. This proves that our error propagation approach is careful and conservative. Values of masses and their precision depend on inclination and its error that are derived from photometry. To decouple our analysis from LC data, we may compare $M_{1,2}\sin^3i$ values obtained from BACHES and HARPS spectra:  $M_1\sin^3i = 1.30 (0.06)$M$_\Sun$, $M_2\sin^3i = 1.61 (0.07)$M$_\Sun$ for BACHES data and $M_1\sin^3i= 1.3367 (0.0009)$M$_\Sun$, $M_2\sin^3i = 1.6218 (0.0015)$M$_\Sun$ for HARPS. Again we see that values differ by 2.7\% and 0.7\%. Following this approach we conclude that we can trust the RV measurements obtained with BACHES and our pipeline and continue the ongoing second survey observing campaign.

Presented plots show that RVs of bright SB2 systems (HD60803 and HD87810, Fig. \ref{fig:plotHD60803}) have residuals that are dominated by systematic errors. This is manifested by equal signs of residuals for both components. To estimate the real dispersion of the RVs free from systematic errors, we computed a pseudo-SB1 model by fitting an orbit to RV$_1$$ - $RV$_2$ values. For HD60803 the resulting solution has $v_\gamma \approx 0$ km s$^{-1}$ (which is expected) and a RMS of 0.31 km s$^{-1}$. Compared to the original RMS values presented in Tab. \ref{tab:SB2}, we observe a twofold drop of RMS value which is a good estimation of the actual performance of the instrument that can be expected in an isolated environment. In case of fainter eclipsing binaries photon noise has a much stronger contribution to the errors so this effect is not present.

Temperatures of the brighter stars of the studied binary systems were obtained from the color-temperature calibration by \cite{Worthey2011} using color values from the \textsc{tycho-2} catalogue \citep{Hog2000}. These temperatures were fixed when calculating the temperatures of the other components of the systems using the \textsc{phoebe} code. Since the tabulated values of temperatures are based on the colors of the whole system (not just a brighter component), the given values are approximated. In order to solve that question multi-color photometry or spectral disentangling is essential, but is beyond the scope of this work.

Future work involves implementing a fully automated data acquisition routine that will increase the efficiency. A temperature vs. spectrum shift model can be derived and used to better calibrate spectra. Especially when the ambient temperature changes abruptly during the exposure, using two calibration ThAr spectra with same weights might be a too simple approach. This, however, requires a precise sensor to be installed in the vicinity of the spectrograph. An active optics system is under development and in the future may allow us to reach $V = 11$ mag stars with a similar SNR. This would increase the amount of targets suitable to observe with the described setup by a factor of 2.5. 

The bottom line is that with our setup, we can spectroscopically characterize about 300 eclipsing binary stars per year up to 10.2 mag assuming typical weather conditions at SAAO. All of this without a single observing trip and soon in a fully automated way that does not require human oversight.

\acknowledgments
We are grateful to the technical and administration staff in SAAO for their help during the commissioning of the instrument.
This work is supported by the European Research Council through a Starting Grant, the National Science Centre through grant 5813/B/H03/2011/40, the Polish Ministry of Science and Higher Education through grant 2072/7.PR/2011/2 and the Foundation for Polish Science through \textit{Idee dla Polski} funding scheme. K.G.H. acknowledges support provided by the National Astronomical Observatory of Japan as Subaru Astronomical Research Fellow. P.S. acknowledges support provided by the National Science Center through grant 2011/03/N/ST9/03192. M.R. acknowledges support provided by the National Science Center through grant 2015/16/S/ST9/00461.

\bibliographystyle{apj}
\bibliography{ms}

\clearpage

\appendix
\section{Single absolute velocity measurements}
\begin{table*}[htp]
\caption{RV measurements for spectroscopic binaries.}
\begin{center}
\scriptsize{
\begin{tabular}{crrrrrrr}
\hline
\hline
	&	MJD	&	RV$_1$	&	$\sigma$RV$_1$	&	O-C	&	RV$_2$	&	$\sigma$RV$_2$	&	O-C	\\ \hline
HD60803	&	57069.8866	&	24.12	&	0.80	&	0.45	&	-14.47	&	0.70	&	0.63	\\
	&	57070.8612	&	28.78	&	0.80	&	-1.00	&	-22.06	&	0.70	&	-0.81	\\
	&	57071.8882	&	35.08	&	0.80	&	-0.49	&	-27.59	&	0.70	&	-0.50	\\
	&	57072.8852	&	40.81	&	0.80	&	0.40	&	-31.82	&	0.70	&	0.16	\\
	&	57075.8462	&	47.92	&	0.80	&	0.14	&	-39.47	&	0.70	&	-0.06	\\
	&	57076.8376	&	45.84	&	0.80	&	-0.72	&	-38.57	&	0.70	&	-0.39	\\
	&	57083.8599	&	-39.69	&	0.80	&	-0.52	&	47.91	&	0.70	&	-0.39	\\
	&	57085.9079	&	-45.83	&	0.80	&	0.93	&	56.58	&	0.70	&	0.63	\\
	&	57086.9071	&	-43.42	&	0.80	&	0.36	&	53.23	&	0.70	&	0.28	\\
	&	57089.8623	&	-24.27	&	0.80	&	-0.48	&	32.92	&	0.70	&	0.15	\\
	&	57090.8878	&	-16.79	&	0.80	&	-1.29	&	23.30	&	0.70	&	-1.11	\\
	&	57091.8418	&	-7.11	&	0.80	&	0.66	&	16.55	&	0.71	&	-0.06	\\
	&	57097.8486	&	35.16	&	0.80	&	0.86	&	-24.77	&	0.70	&	1.04	\\
	&	57099.9122	&	44.40	&	0.80	&	0.69	&	-34.86	&	0.70	&	0.43	\\
	&		&		&		&		&		&		&		\\
HD87810	&	57065.9295	&	-35.43	&	1.23	&	1.07	&	36.82	&	1.23	&	1.33	\\
	&	57067.9020	&	-20.83	&	1.22	&	2.19	&	24.08	&	1.22	&	2.09	\\
	&	57070.9238	&	10.98	&	1.22	&	-0.62	&	-13.41	&	1.22	&	-0.73	\\
	&	57071.9232	&	28.51	&	1.22	&	-0.88	&	-31.38	&	1.22	&	-0.89	\\
	&	57072.9450	&	53.75	&	1.22	&	1.01	&	-52.67	&	1.22	&	1.20	\\
	&	57075.8591	&	-19.51	&	1.22	&	-0.20	&	18.15	&	1.22	&	-0.11	\\
	&	57076.8946	&	-39.02	&	1.22	&	-1.17	&	35.47	&	1.22	&	-1.36	\\
	&	57083.9192	&	12.42	&	1.22	&	0.09	&	-13.46	&	1.22	&	-0.04	\\
	&	57086.8982	&	71.05	&	1.22	&	0.05	&	-72.27	&	1.22	&	-0.13	\\
	&	57089.9205	&	-38.73	&	1.22	&	-0.38	&	36.85	&	1.22	&	-0.49	\\
	&	57091.9012	&	-35.99	&	1.23	&	0.14	&	35.39	&	1.23	&	0.28	\\
	&	57099.9272	&	71.96	&	1.22	&	1.15	&	-70.71	&	1.22	&	1.23	\\
	&	57100.8225	&	31.04	&	1.22	&	-0.92	&	-33.98	&	1.22	&	-0.90	\\
	&	57110.7975	&	28.47	&	1.22	&	-1.51	&	-32.53	&	1.22	&	-1.45	\\
\hline
\end{tabular}
}
\end{center}
\label{tab:ModelsHD}
\end{table*}

\begin{table*}[htp]
\caption{RV measurements for spectroscopic standards.}
\begin{center}
\scriptsize{
\begin{tabular}{crrrrrrr}
\hline
\hline
	&	MJD	&	RV	&	$\sigma$RV	\\ \hline
HD45184	&	2457066.3462	&	-5.26	&	0.05	\\	
	&	2457068.3263	&	-3.78	&	0.05	\\	
	&	2457070.3297	&	-4.63	&	0.05	\\	
	&	2457071.3273	&	-6.78	&	0.04	\\	
	&	2457072.3261	&	-5.91	&	0.05	\\	
	&	2457073.3295	&	-4.33	&	0.05	\\	
	&	2457076.2886	&	-5.49	&	0.05	\\	
	&	2457077.3040	&	-6.30	&	0.04	\\	
	&	2457078.3172	&	-5.02	&	0.05	\\	
	&	2457079.3676	&	-5.43	&	0.05	\\	
	&	2457084.3020	&	-6.93	&	0.06	\\	
	&	2457086.3745	&	-4.35	&	0.06	\\	
	&	2457090.3071	&	-6.07	&	0.05	\\	
	&	2457091.3720	&	-7.34	&	0.06	\\	
	&	2457092.2842	&	-5.71	&	0.05	\\	
	&	2457101.2932	&	-6.04	&	0.05	\\	
	&		&		&		\\	
HD102365	&	2457068.4296	&	19.36	&	0.05	\\	
	&	2457070.4170	&	17.31	&	0.05	\\	
	&	2457071.4057	&	14.68	&	0.04	\\	
	&	2457072.4110	&	15.24	&	0.04	\\	
	&	2457073.4346	&	17.35	&	0.05	\\	
	&	2457076.3976	&	16.13	&	0.05	\\	
	&	2457077.3831	&	15.45	&	0.04	\\	
	&	2457084.4088	&	16.11	&	0.04	\\	
	&	2457086.4585	&	18.40	&	0.05	\\	
	&	2457087.3729	&	16.02	&	0.04	\\	
	&	2457090.4107	&	15.47	&	0.05	\\	
	&	2457092.3908	&	16.67	&	0.05	\\	
	&	2457101.3632	&	15.36	&	0.04	\\	
	&	2457111.3473	&	15.65	&	0.04	\\	\hline
\hline
\end{tabular}
}
\end{center}
\end{table*}

\begin{table*}[htp]
\caption{Single, absolute RV measurements for eclipsing binaries.}
\begin{center}
\scriptsize{
\begin{tabular}{crrrrrrr}
\hline
\hline
	&	MJD	&	RV$_1$	&	$\sigma$RV$_1$	&	O-C	&	RV$_2$	&	$\sigma$RV$_2$	&	O-C	\\ \hline
J042724-2756.2	&	57067.7862	&	56.09	&	1.92	&	1.38	&	-45.92	&	1.05	&	0.70	\\
	&	57068.7774	&	78.11	&	2.55	&	2.13	&	-64.90	&	1.73	&	1.60	\\
	&	57069.7866	&	64.33	&	3.74	&	1.83	&	-52.32	&	2.61	&	1.57	\\
	&	57071.7851	&	-32.14	&	5.21	&	-2.19	&	31.30	&	2.04	&	-1.27	\\
	&	57072.7877	&	-62.83	&	5.01	&	2.55	&	66.51	&	1.92	&	0.78	\\
	&	57074.7822	&	-33.48	&	6.56	&	5.41	&	47.85	&	5.23	&	6.92	\\
	&	57076.7861	&	53.62	&	2.44	&	-2.52	&	-51.66	&	1.58	&	-3.71	\\
	&	57077.7758	&	75.86	&	4.79	&	-0.27	&	-64.99	&	2.36	&	1.64	\\
	&	57083.7601	&	-43.67	&	3.57	&	-5.54	&	39.87	&	2.31	&	-0.36	\\
	&	57086.7592	&	74.12	&	2.65	&	-2.07	&	-67.10	&	1.53	&	-0.40	\\
	&	57089.7647	&	-31.54	&	2.98	&	1.12	&	36.00	&	1.44	&	0.89	\\
	&	57090.8258	&	-67.60	&	4.63	&	0.00	&	65.91	&	2.51	&	-1.91	\\
	&	57091.7643	&	-54.09	&	6.77	&	13.82	&	69.50	&	2.37	&	1.40	\\
	&	57097.7590	&	15.07	&	6.40	&	-0.88	&	-15.75	&	4.76	&	-5.37	\\
	&	57100.7523	&	-68.95	&	3.62	&	-1.51	&	67.34	&	2.25	&	-0.32	\\
	&	57110.7525	&	-34.35	&	3.33	&	-1.14	&	34.57	&	2.30	&	-1.05	\\
	&		&		&		&		&		&		&		\\
J061212-1215.8	&	57065.8305	&	100.88	&	3.05	&	2.83	&	-113.44	&	3.53	&	-2.35	\\
	&	57067.8112	&	-61.81	&	3.09	&	1.56	&	27.60	&	3.53	&	2.36	\\
	&	57069.8133	&	-89.66	&	3.15	&	-1.97	&	39.14	&	3.59	&	-6.66	\\
	&	57071.8106	&	89.58	&	3.18	&	-5.51	&	-105.58	&	3.56	&	3.00	\\
	&	57072.8130	&	-98.58	&	3.14	&	-1.63	&	49.38	&	3.60	&	-4.26	\\
	&	57074.8074	&	95.34	&	3.27	&	3.63	&	-100.02	&	3.66	&	5.71	\\
	&	57075.7737	&	-103.03	&	3.18	&	-3.90	&	57.64	&	3.71	&	2.18	\\
	&	57077.8016	&	93.43	&	3.21	&	6.01	&	-103.47	&	3.58	&	-1.36	\\
	&	57083.7862	&	74.29	&	3.27	&	-2.11	&	-94.47	&	3.61	&	-1.65	\\
	&	57084.7628	&	-117.66	&	3.09	&	0.96	&	72.34	&	3.54	&	0.39	\\
	&	57086.7838	&	68.75	&	3.07	&	0.00	&	-88.15	&	3.54	&	-1.80	\\
	&	57089.7915	&	55.43	&	3.51	&	-2.87	&	-77.57	&	3.54	&	-0.03	\\
	&	57090.8508	&	-127.21	&	3.35	&	1.71	&	85.40	&	3.58	&	4.72	\\
	&	57099.8481	&	-119.83	&	3.19	&	3.68	&	76.97	&	3.55	&	0.88	\\
	&	57100.7765	&	56.70	&	3.12	&	-2.73	&	-79.33	&	3.61	&	-0.84	\\
	&		&		&		&		&		&		&		\\
J071626+0548.8	&	57065.8674	&	35.64	&	1.41	&	1.67	&	-5.39	&	1.95	&	2.24	\\
	&	57067.8457	&	81.70	&	1.46	&	1.01	&	-54.64	&	1.80	&	0.09	\\
	&	57069.8477	&	33.91	&	1.45	&	0.76	&	-5.46	&	1.75	&	1.36	\\
	&	57071.8468	&	-35.17	&	1.52	&	-0.49	&	60.04	&	1.81	&	-1.56	\\
	&	57072.8471	&	-40.90	&	1.45	&	-0.81	&	66.60	&	1.72	&	-0.47	\\
	&	57075.8067	&	-2.33	&	2.89	&	1.50	&	31.79	&	1.77	&	1.31	\\
	&	57077.8368	&	43.24	&	1.70	&	-2.65	&	-22.72	&	2.12	&	-3.06	\\
	&	57078.8849	&	71.31	&	1.40	&	-1.16	&	-46.64	&	1.68	&	-0.19	\\
	&	57085.8921	&	-29.24	&	1.40	&	-0.90	&	55.08	&	1.69	&	-0.13	\\
	&	57086.8106	&	-14.07	&	1.43	&	-0.15	&	38.62	&	1.75	&	-2.04	\\
	&	57089.8242	&	58.17	&	1.41	&	-0.10	&	-32.66	&	1.81	&	-0.53	\\
	&	57091.8017	&	77.07	&	1.46	&	-0.17	&	-51.44	&	1.76	&	-0.17	\\
	&	57097.8101	&	-20.59	&	1.53	&	2.15	&	52.07	&	1.74	&	2.50	\\
		&		&		&		&		&		&		&		\\
	J072222-1159.8	&	57065.8917	&	-49.27	&	1.02	&	0.77	&	100.12	&	1.50	&	0.42	\\
	&	57067.8683	&	-40.82	&	1.18	&	0.38	&	90.43	&	1.85	&	0.78	\\
	&	57069.8707	&	39.52	&	1.19	&	-0.19	&	-2.56	&	2.03	&	-0.48	\\
	&	57070.8451	&	56.19	&	1.23	&	-2.18	&	-25.97	&	1.47	&	-2.74	\\
	&	57071.8704	&	67.72	&	1.71	&	0.06	&	-34.46	&	1.96	&	-0.71	\\
	&	57072.8689	&	69.03	&	1.31	&	-0.56	&	-32.88	&	1.95	&	3.07	\\
	&	57075.8292	&	43.29	&	1.71	&	0.16	&	-0.56	&	6.33	&	5.39	\\

\hline
\end{tabular}
}
\end{center}
\label{tab:ModelsDEB}
\end{table*}
\addtocounter{table}{-1}
\begin{table*}[htp]
\caption{(continued)}
\begin{center}
\scriptsize{
\begin{tabular}{crrrrrrr}
\hline
\hline
	&	MJD	&	RV$_1$	&	$\sigma$RV$_1$	&	O-C	&	RV$_2$	&	$\sigma$RV$_2$	&	O-C	\\ \hline
	&	57077.8604	&	-13.16	&	2.95	&	-2.73	&	50.61	&	4.22	&	-4.16	\\
	&	57078.9067	&	-50.63	&	1.16	&	-0.83	&	100.35	&	1.94	&	0.94	\\
	&	57083.8432	&	56.31	&	2.07	&	-1.74	&	-24.05	&	2.67	&	-1.19	\\
	&	57086.8326	&	66.99	&	0.91	&	0.66	&	-31.20	&	1.66	&	1.04	\\
	&	57091.8248	&	-46.09	&	1.83	&	-0.06	&	93.46	&	2.08	&	-1.68	\\
	&	57097.8330	&	70.00	&	2.45	&	2.78	&	-38.69	&	6.62	&	-5.43	\\
	&	57099.8964	&	67.49	&	1.24	&	1.41	&	-30.16	&	1.91	&	1.80	\\
		&		&		&		&		&		&		&		\\
J111134-4956.2	&	57075.9138	&	1.23	&	2.97	&	1.02	&	18.44	&	2.11	&	3.19	\\
	&	57076.8685	&	-42.02	&	1.99	&	-2.44	&	58.36	&	1.73	&	-1.82	\\
	&	57078.9292	&	-49.92	&	3.74	&	-1.00	&	71.41	&	2.78	&	0.69	\\
	&	57083.8940	&	-12.29	&	2.29	&	-2.06	&	24.77	&	2.02	&	-2.27	\\
	&	57085.9432	&	-70.75	&	1.51	&	0.89	&	98.13	&	1.43	&	1.73	\\
	&	57086.8559	&	-33.77	&	2.40	&	1.73	&	54.94	&	2.00	&	-0.62	\\
	&	57089.8952	&	65.09	&	3.98	&	5.68	&	-51.29	&	2.99	&	0.25	\\
	&	57090.9189	&	17.87	&	3.65	&	-2.14	&	-6.67	&	2.57	&	0.41	\\
	&	57091.8732	&	-18.08	&	3.49	&	2.45	&	38.56	&	3.11	&	-0.12	\\
	&	57099.9728	&	-32.17	&	2.47	&	3.33	&	56.52	&	1.63	&	0.95	\\
	&	57100.8476	&	-67.84	&	2.85	&	-2.89	&	86.71	&	1.79	&	-2.12	\\
	&	57110.8310	&	31.65	&	2.30	&	-1.68	&	-22.50	&	1.77	&	-0.35	\\
	&		&		&		&		&		&		&		\\
J155259-6637.8	&	57085.9696	&	-61.18	&	2.01	&	0.42	&	99.11	&	1.75	&	0.51	\\
	&	57100.8907	&	52.95	&	2.03	&	-1.93	&	-47.17	&	2.05	&	-1.86	\\
	&	57110.8596	&	67.09	&	2.03	&	-1.29	&	-64.67	&	1.83	&	-2.68	\\
	&	57120.8952	&	-40.24	&	2.03	&	-1.00	&	70.19	&	1.81	&	-0.77	\\
	&	57129.8414	&	39.65	&	2.05	&	3.45	&	-20.79	&	2.02	&	1.45	\\
	&	57133.9572	&	74.81	&	2.01	&	-0.44	&	-68.93	&	1.75	&	1.54	\\
	&	57134.8833	&	80.25	&	2.01	&	1.72	&	-73.83	&	1.74	&	0.67	\\
	&	57146.8682	&	49.61	&	2.02	&	-0.87	&	-38.91	&	2.02	&	0.96	\\
		&		&		&		&		&		&		&		\\
	J155259-6637.8	&	55720.9595	&	83.77959	&	0.08035	&	0.05378	&	-77.82630	&	0.15044	&	-0.17658	\\
HARPS	&	55720.9873	&	84.53926	&	0.08051	&	0.07291	&	-78.68926	&	0.15051	&	-0.14198	\\
	&	55721.0909	&	86.64920	&	0.08034	&	0.02615	&	-81.19174	&	0.15017	&	-0.03053	\\
	&	55721.1988	&	87.83279	&	0.08048	&	-0.00515	&	-82.55724	&	0.15064	&	0.07640	\\
	&	55722.0164	&	63.54339	&	0.08035	&	-0.03133	&	-53.37987	&	0.15017	&	-0.15773	\\
	&	55722.0304	&	62.79337	&	0.08032	&	0.08667	&	-52.39798	&	0.15027	&	-0.22822	\\
	&	55722.1077	&	57.50520	&	0.08027	&	-0.18795	&	-45.86274	&	0.15022	&	0.22827	\\
	&	55722.2301	&	49.04773	&	0.08021	&	-0.04515	&	-35.52648	&	0.15016	&	0.13600	\\
	&	55722.2970	&	44.12452	&	0.08043	&	0.03633	&	-29.55767	&	0.15023	&	0.03559	\\
	&	55810.9929	&	-45.71668	&	0.08074	&	-0.01382	&	79.19445	&	0.15058	&	-0.17862	\\
	&	55811.0045	&	-45.01693	&	0.08031	&	0.01880	&	78.32301	&	0.15045	&	-0.23993	\\
	&	55811.0654	&	-41.51768	&	0.08057	&	-0.13909	&	74.29274	&	0.15030	&	0.17072	\\
	&	55811.9789	&	30.28394	&	0.08047	&	0.02929	&	-12.70619	&	0.15023	&	0.10873	\\
	&	55812.0375	&	34.90990	&	0.08058	&	-0.08714	&	-18.39106	&	0.15047	&	0.17617	\\
	&	55812.0877	&	38.94375	&	0.08076	&	-0.03620	&	-23.25676	&	0.15035	&	0.14123	\\
	&	55813.0465	&	87.23705	&	0.08035	&	0.04707	&	-81.89259	&	0.15031	&	-0.04426	\\
	&	56136.9868	&	-40.05083	&	0.08108	&	-0.01977	&	72.68255	&	0.15084	&	0.19679	\\
	&	56137.0600	&	-44.35846	&	0.08065	&	0.13422	&	77.75319	&	0.15028	&	-0.15029	\\
	&	56137.1279	&	-48.46208	&	0.08049	&	-0.14708	&	82.71919	&	0.15021	&	0.17399	\\
	&	56137.1970	&	-51.82894	&	0.08026	&	0.04235	&	86.69130	&	0.15029	&	-0.17279	\\
	&	56137.9697	&	-64.23682	&	0.08040	&	-0.04757	&	101.83131	&	0.15020	&	0.00608	\\
	&	56138.0530	&	-62.25392	&	0.08052	&	0.04902	&	99.55356	&	0.15033	&	0.01957	\\
	&	56138.1280	&	-60.01159	&	0.08059	&	0.07293	&	96.73654	&	0.15043	&	-0.10288	\\
	&	56138.2123	&	-56.95636	&	0.08035	&	0.06838	&	93.14042	&	0.15029	&	0.01739	\\
	&	56178.9792	&	-24.28673	&	0.08059	&	-0.05603	&	53.39383	&	0.15046	&	0.09155	\\
	&	56179.0811	&	-16.30516	&	0.08137	&	0.08153	&	43.82394	&	0.15064	&	0.04354	\\
	
\hline
\end{tabular}
}
\end{center}
\end{table*}

\end{document}